\documentclass[onecolumn,draftcls]{IEEEtran}
\usepackage[noadjust]{cite}

\ifCLASSINFOpdf
  
\else
\usepackage[dvips]{graphicx,color}
  
\fi
\usepackage[cmex10]{amsmath}

\usepackage{multirow}
\usepackage{amsmath,amssymb,mathrsfs}
\usepackage{type1cm}

\newtheorem{teiri}{Theorem}
\newtheorem{kei}{Corollary}
\newtheorem{hodai}{Lemma}
\newtheorem{teigi}{Definition}
\newtheorem{rei}{Example}
\newtheorem{chui}{Remark}

\newcommand{\bal}{\begin{align}}
\newcommand{\non}{\nonumber}

\renewcommand{\IEEEQED}{\hfill\IEEEQEDopen}
\newcommand{\Exp}{{\rm E}}

\newcommand{\spr}[1]{{\bf #1}}

\newcommand{\vep}{\varepsilon}
\newcommand{\vph}{\varphi}

\renewcommand{\thesection}{\Roman{section}}

\newcommand{\cB}{{\cal B}}

\newcommand{\cH}{{\cal H}}
\newcommand{\cM}{{\cal M}}

\newcommand{\cS}{{\cal S}}
\newcommand{\cU}{{\cal U}} 

\newcommand{\cX}{{\cal X}}
\newcommand{\cY}{{\cal Y}}

\newcommand{\cZ}{{\cal Z}}

\newcommand{\sC}{\spr{C}}

\newcommand{\sS}{\spr{S}}
\newcommand{\sT}{\spr{T}}
\newcommand{\sU}{\spr{U}}

\newcommand{\sX}{\spr{X}}
\newcommand{\sY}{\spr{Y}}
\newcommand{\sZ}{\spr{Z}}

\newcommand{\ssc}{\spr{c}}
\newcommand{\ssu}{\spr{u}}
\newcommand{\ssx}{\spr{x}}
\newcommand{\ssz}{\spr{z}}
\newcommand{\ssy}{\spr{y}}
\newcommand{\sss}{\spr{s}}

\newcommand{\sst}{\spr{t}}

\newcommand{\nth}{\frac{1}{n}}

\newcommand{\bteiri}{\begin{teiri}}
\newcommand{\eteiri}{\end{teiri}}
\newcommand{\bkei}{\begin{kei}}
\newcommand{\ekei}{\end{kei}}
\newcommand{\brei}{\begin{rei}}
\newcommand{\erei}{\end{rei}}
\newcommand{\bhodai}{\begin{hodai}}
\newcommand{\ehodai}{\end{hodai}}
\newcommand{\bteigi}{\begin{teigi}}
\newcommand{\eteigi}{\end{teigi}}
\newcommand{\bchui}{\begin{chui}}
\newcommand{\echui}{\end{chui}}
\newcommand{\beq}{\begin{equation}}
\newcommand{\eeq}{\end{equation}}
\newcommand{\beqn}{\begin{eqnarray}}
\newcommand{\eeqn}{\end{eqnarray}}
\newcommand{\beqns}{\begin{eqnarray*}}
\newcommand{\eeqns}{\end{eqnarray*}}

\newcommand{\map}{\vph_n: \cX^n \to \cY^n}
\newcommand{\mapMtoY}{\vph_n: \cM_{M_n} \to \cY^n}
\newcommand{\mapXtoM}{\vph_n: \cX^n \to \cM_{M_n}}

\begin{document}
%
\title{
Wiretap Channels with \\  Causal State Information: Strong Secrecy
}
\author{
Te~Sun~Han,~\IEEEmembership{Life Fellow,~IEEE},
\thanks{T. S. Han is  with the
Quantum ICT Advanced Development Center, National Institute of Information and
Communications Technology (NICT), Nukui-kitamachi 4-2-1, Koganei,
Tokyo,184-8795, Japan (email: han@is.uec.ac.jp)} 
Masahide Sasaki\thanks{M. Sasaki is  with the
Advanced  ICT Research Institute, NICT, Nukui-kitamachi 4-2-1, Koganei,
Tokyo,184-8795, Japan (email: psasaki@nict.go.jp)}
\  \ \mbox{}\thanks{Copyright (c) 2017 IEEE. Personal use of this material is permitted.  
However, permission to use this material for any other purposes must be obtained from the IEEE by sending 
a request to pubs-permissions@ieee.org}}
%

%
\maketitle
\begin{abstract} 
 The coding problem for wiretap channels with  {\em causal} channel state information available at the encoder and/or the decoder
is studied under the {\em strong secrecy} criterion. This problem consists of two aspects: one is due to 
 wiretap channel coding and the other is due to one-time pad cipher based on the secret key agreement
between Alice and Bob using the channel state information. These two aspects are closely related to each other 
and 
give rise to an intriguing tradeoff between exploiting the state to  boost secret-message rates versus extracting cryptographic key to improve secrecy capabilities. This issue has yet to be understood how to optimally
reconcile the two. We newly devised the “iterative” forward-backward coding scheme, combining wiretap channel coding and secret-key-agreement-based one-time pad cipher. We then established reasonable lower bounds of the secrecy capacity for wiretap channels with causal channel state information available only at the encoder (Theorem 1), which can be easily extended to general cases with various kinds of correlated channel state information at the encoder (Alice), decoder (Bob) and wiretapper (Eve).  In particular, for degraded wiretap channels, we give the secret-message 
(secret-key) capacity bounds (Theorems \ref{teiri:RR-1}, \ref{teiri:RRR1}). 
 
\end{abstract}


\begin{IEEEkeywords}
wiretap  channel, channel state information,  causal coding, secret key agreement, secrecy capacity,
strong secrecy
 \end{IEEEkeywords}

%

\section{Introduction}\label{introduction1}

In this paper the coding problem for the wiretap channel (WC) with {\em causal} channel state information (CSI) available at the encoder (Alice) and/or the decoder (Bob)
is studied. The concept of WC ({\em without} CSI) originates in Wyner \cite{wyner-wire} and was extended to a more general WC by Csisz\'ar and K\"orner \cite{csis-kor-3rd}. 
These landmark papers have been followed by many subsequent extensions and generalizations from the 
viewpoint of theory and practice.
In particular, among others,  the 
WC {\em with} CSI  has also been extensively investigated in the literature.
Early works include 
Mitrpant, Vinck and Luo \cite{mitrpant},  Chen and Vinck \cite{chen-vinck},
and  Liu and Chen \cite{liu-chen}  that have studied  
the {\em capacity-equivocation} region for degraded WCs with {\em non-causal} CSI to 
establish inner and/or outer bounds on the region, which was motivated  by physical-layer 
security problems to actually intervene  in practical fading channel communications.
Moreover, subsequent recent developments in this direction with {\em non-causal} CSI can be found also in 
Dai, Zhuang and Vinck \cite{dai-vinck}, Boche and Schaefer \cite{boche}, Dai and Luo \cite{dai-luo},
  Prabhakaran {\em et al.} 
   \cite{prabhakaran},  Goldfeld 
  {\em et al.} \cite{goldfeld}, Bunin {\em et al.} \cite{bunin},
etc.

Generally speaking,  the coding  scheme with causal/non-causal CSI outperforms 
 the one without CSI, because
knowledge of the CSI enables us to share a common secret key between Alice and Bob
to augment the secrecy capacity.  More specifically, then,
in addition to  the standard WC coding  (called the {\em Wyner's 
 WC coding} \cite{wyner-wire}, \cite{csis-kor-3rd}) 
  {\em without} resorting to the CSI,
we may incorporate also
 the cryptographic  scheme called 
  the {\em Shannon's one-time pad (OTP)  cipher}
 (cf. Shannon \cite{shannon-secrecy})
 based on the {\em secret key agreement} 
 (cf. Maurer \cite{maurer}, Ahlswede and Csisz\'ar \cite{ahls-csis}) 
 using the CSI between Alice and Bob.
Thus, the problem consists of two aspects: one is due to  wiretap channel coding and the other is due to one-time pad cipher based on the secret key agreement. 
Here is the trade-off between them depending on  how to use the state information $S$.

 Recent works taking account of such a secrecy key agreement aspect include Khisti, Diggavi and Wornell [14],
Chia and El Gamal [17], Sonee and Hodtani [19], and Fujita [20]. In particular,  [14] addresses the problem of key capacity that focuses on the maximum rate of secret key agreement between Alice and Bob rather than on the maximum rate of secure message transmission. However, we cannot say that the secrecy capacity problem in these works with causal CSI has now been fully solved. This is because the problem with causal/non-causal CSI necessarily includes the two separate but closely related coding schemes as mentioned in the above paragraph.  

Among others,   Chia and El Gamal \cite{chia-elgamal} addresses
 the case with {\em causal} common CSI available  at both 
 Alice and Bob, whereas Fujita \cite{fujita} deals with the case with {\em causal} CSI available only at Alice (given a {\em physically degraded} WC).
Both includes lower bounds on the {\em weak} secrecy capacity, but with tight secrecy capacity formulas
in special cases. The present paper is motivated mainly by these two papers, and
the main result to be given in this paper is in nice accordance with  their results.
In particular, we have newly established the “iterative” forward-backward coding scheme for WCs with causal CSI available at Alice with reasonable lower bounds on secrecy capacity. For degraded channels, we successfully established not only lower/upper bounds, but also several exact secret-massage (secret-key) capacities. 

%
%


%
The present paper is organized as follows.

In Section \ref{intro-ge-cs0}, we give the statement of the problem and the key result 
(Theorem \ref{teiri:main1}) for the WC with {\em causal} CSI 
available only at Alice
along with comparison with  the work of  Chia and El Gamal \cite{chia-elgamal}.

In Section \ref{intro-ge-cs1}, we give the detailed  proof of Theorem \ref{teiri:main1}
to establish lower bounds on the {\em strong} secrecy capacity.
The main ingredients for the proof are Slepian-Wolf coding, Csisz\'ar-K\"orner's key construction, 
Gallager's maximum likelihood decoding, and 
Han-Verd\'u's resolvability argument, where
in the process of these proofs we do {\em not} invoke the argument of typical sequences at all,
which enables us to cope with alphabets that are not necessarily finite
(e.g., for  Gaussian WCs).
  
%

In Section \ref{sec:illust}, in order to obtain insights into the significance of Theorem \ref{teiri:main1},
we provide  specific secrecy capacity bounds (including  upper/lower bounds)
for degraded WCs with causal/non-causal CSI (Theorems \ref{teiri:RR-1}, \ref{teiri:RRR1}, \ref{teiri:Naga1} 
and Corollaries
\ref{kei:RR1}, \ref{kei:RR2}, \ref{kei:han-sasaki}).

In Section \ref{sec:previous}, 
since the present work has partly close bearing with that of Fujita \cite{fujita},
we compare both of them to scrutinize the details of these works. 


%

In Section \ref{conc-remark}, we conclude the paper with several remarks.

\section{Problem Statement and the Result}\label{intro-ge-cs0}

\begin{figure}[htbp]
\begin{center}
\includegraphics[width=80mm]{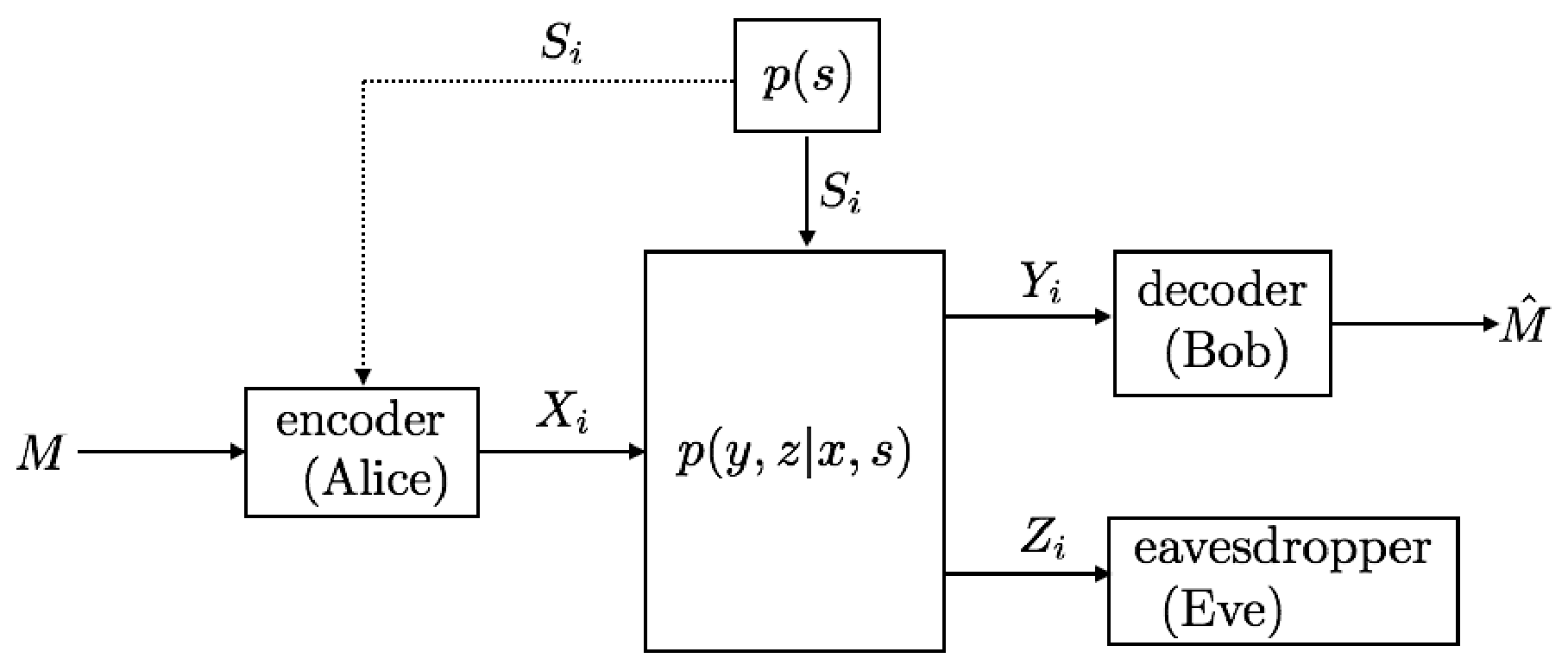}
\end{center}
\caption{WC with CSI available only at Alice ($i=1,2,\cdots, n$).
          }
\label{fig-c2}
\end{figure}

%
%
A {\em stationary memoryless} WC 
as illustrated in Fig. \ref{fig-c2} is specified by giving the   conditional (transition) probability
\beq\label{eq:causal1}
p(y,z|x,s) =P_{YZ|XS}(y,z|x,s)
\eeq 
with  input random variable $X$ (for Alice),
 outputs random variables $Y$ (for Bob), $Z$ (for Eve), and  CSI random variable
$S$, which are assumed to take values in
 alphabets $\cX, \cY, \cZ, \cS$, respectively.
Alice $X$ (sender),
who only has access to {\em stationary memoryless}
CSI $S$ available,
 wants to send a confidential message $M\in \cM=[1: 2^{nR}]$ (over  $n$  channel
transmissions)
to Bob $Y$ (legitimate receiver) 
while keeping it secret from Eve $Z$ (eavesdropper),  where we use here and hereafter  the notation 
$[i:j]=\{i, i+1,\cdots, j-1, j\}$ for $j\ge i$, and $R\ge 0$ is called the {\em rate}.

An $(n, 2^{nR})$ code for the WC with {\em causal} CSI $S$ at the encoder 
consists of \\
 (i) a message set $\cM=[1: 2^{nR}]$, \\
 (ii) a {\em stochastic} ``causal"  encoder  
$f_i: \cM\times \cS^i\to \cX$ subject to conditional probability
$p(x| m, s^i)$\\
\quad\   to produce  the channel input $X_i(M) =f_i(M, S^i)$
 at each  time  $i\in [1:n]$, and \\
 (iii) a decoder $g: \cY^n \to \cM$ 
  (for Bob) to assign an estimate $\hat{M}$
to each received sequence  $\sY$, 
where we use the notation $a^i=a_1a_2
\cdots a_i$ (in particular,   ${\bf a}= a_1a_2\cdots a_n$: the bold-faced letters indicate 
sequences of length $n$) and assume that the message $M$ is {\em uniformly} distributed on the message set $\cM$. 

The probability of error is defined to be $P_e=\Pr\{\hat{M}\neq M\}$.
The {\em information leakage} at Eve with output sequence $\sZ$,
which measures the amount of information about $M$ that leaks out to Eve, is defined to be
$I_E=I(M;\sZ)$ (the mutual information between $M$ and $\sZ$). 
%
 It should be noted here that this measure is {\em not}
$R_E=\nth I(M;\sZ)$
(the information leakage {\em rate}). This means that in this paper we are concerned only with the {\em strong secrecy}
but not the
{\em weak secrecy}  as was the case in the literature (e.g., cf. Chia and El Gamal \cite{chia-elgamal},
Fujita \cite{fujita}).

A secrecy rate $R$ is said to be achievable if there exists a sequence of codes $(n, 2^{nR})$ with
$P_e\to 0$ and $I_E\to 0$ as $n\to \infty$. The  {\em secrecy capacity} 
with CSI available only at the encoder (=E), denoted by 
$C_{\mbox{{\scriptsize\rm CSI-E}}}$, is the supremum of all achievable rates.

In order to implement the coding scheme for the WC, it is convenient to introduce its associated
 channel $\omega$ as follows: Let $U$ be an arbitrary auxiliary random variable with values in a set $\cU$
that is {\em independent} of the CSI  variable $S$, and let $h: \cU\times \cS \to \cX$ be a {\em stochastic} mapping subject to conditional probability $p(x|u, s)$. According to the Shannon strategy \cite{shannon-cent}, we
  define the $\omega$ as the WC
specified by the conditional probability
 \beq\label{eq:causal2} 
p(y,z|u, s) =\sum_{x\in \cX}p(y,z|x,s)p(x|u, s),
\eeq 
which gives the associated  WC (called a {\em test channel}) with input variable $U$ (Alice), outputs variables $Y, Z$ 
(Bob and Eve)
and CSI variable $S$. 
Thus, hereafter we may focus solely on the coding problem for the channel $\omega$
from the standpoint of achievabie rates.

%
\smallskip
Let us now describe the main result. Set 
\beqn
R_{\mbox{{\scriptsize\rm CSI-0}}}(p(u), p(x|u,s)) &=& I(U;Y)-I(U;Z),\label{eq:naosi-1}\\
R_{\mbox{{\scriptsize CSI-1}}}(p(u), p(x|u,s))
& = & \min \Bigl[I(U; Y) -I(U; SZ)\nonumber\\
& & \qquad + H(S|Z)-H(S|UY), \nonumber\\
& & \quad\qquad\quad I(U; Y)-H(S|UY)\Bigr],\label{eq:rata-2}\\
R_{\mbox{{\scriptsize CSI-2}}}(p(u), p(x|u,s))
&=& \min \Bigl[H(S|UZ)-H(S|UY), I(U;Y)-H(S|UY)\Bigr]\label{eq:rata-3},
\eeqn
where 
$I(\cdot; \cdot), I(\cdot; \cdot|\cdot)$ denote the (conditional) mutual informations; 
and $H(\cdot), H(\cdot|\cdot)$ denote the (conditional) entropies.
Moreover, for simplicity we use the notation $A_1A_2\cdots A_m$ to denote $(A_1,A_2,\cdots, A_m)$.

Then, we have
the following lower bound on the 
{\em secrecy capacity} 
 $C_{\mbox{{\scriptsize\rm CSI-E}}}$
 with the understanding that $R_{\mbox{{\scriptsize CSI-1}}}(p(u),$ $p(x|u,s))=0$
 when $I(U; Y) -I(U; SZ)<0$ or $H(S|Z)-H(S|UY)<0$:

\smallskip
\bteiri\label{teiri:main1}
\mbox{} Let us consider the WC with CSI as in Fig.\ref{fig-c2}
with {\em causal }CSI available only at Alice.
Then, the  secrecy capacity $C_{\mbox{{\scriptsize\rm CSI-E}}}$ is lower bounded as 
\beqn\label{eq:sed1}
C_{\mbox{{\scriptsize\rm CSI-E}}} &\ge&  \max \Bigl[ \ 
\max_{p(u), p(x|u,s)}
R_{\mbox{{\scriptsize\rm CSI-0}}}(p(u), p(x|u,s)),\nonumber\\
& & \quad\qquad \max_{p(u), p(x|u,s)}
R_{\mbox{{\scriptsize\rm CSI-1}}}(p(u), p(x|u,s)),\nonumber\\
& &\quad \qquad  \max_{p(u), p(x|u,s)}
R_{\mbox{{\scriptsize\rm CSI-2}}}(p(u), p(x|u,s))\Bigr],
\eeqn
\eteiri
where $p(u), p(x|u,s)$ ranges over all possible (conditional) probability 
distributions such that $p(u, s) =p(u)p(s)$, and 
notice here that $p(s)$ is a given distribution and so cannot be varied. \IEEEQED

The term $H(S|UY)$ in  (\ref{eq:rata-2}), (\ref{eq:rata-3})
specifies the rate of (auxiliary)  Slepian-Wolf coding
for information reconcillation in secret key agreement (for OPT cipher) between Alice and Bob using the CSI;
 in (\ref{eq:rata-2}) the term $I(U;Y)-I(U;SZ)$
 specifies the transmission rate of confidential message via 
 WC coding
\footnote[2]{
%
 Notice here that the WC  $\omega$ in this paper is equipped with no public authenticated noiseless channel 
 between Alice and Bob unlike in the standard setting of secret key agreement,  but all communications occur inside the WC $\omega$ in  one-way fashion from Alice to Bob.}; 
%
 the term $H(S|Z)-H(S|UY)$ in  
 (\ref{eq:rata-2})
specifies the key rate to transmit an additional  
confidential message via OTP cipher with the  secret  key shared between Alice and Bob  using the CSI;
 %
 the term $I(U;Y)-H(S|UY)$ in  (\ref{eq:rata-2}),
(\ref{eq:rata-3})
specifies the  upper bound on  total transmission rates for  two kinds of confidential  messages
as above, 
excluding the Slepian-Wolf auxiliary message.

The achievability of 
$R_{\mbox{{\scriptsize\rm CSI-0}}}(p(u), p(x|u,s))$ is well known, which is attained by
 the standard WC coding
{\em without} resorting to the OTP cipher using the  secret key  generated by  CSI (cf. 
 Csisz\'ar and K\"orner \cite{csis-kor-3rd}, El Gamal and Kim \cite{gamal-kim}, Dai and Luo \cite{dai-luo}).
 This is actually attained by employing the ``one-time" CSI coding in the sense of Han, Endo and Sasaki \cite{han-et}.

The  achievability proof for  $R_{\mbox{{\scriptsize\rm CSI-1}}}(p(u), p(x|u,s))$ and 
$R_{\mbox{{\scriptsize\rm CSI-2}}}(p(u), p(x|u,s))$ 
in Theorem \ref{teiri:main1} is provided in the next section. 

\begin{figure}[htbp]
\begin{center}
\includegraphics[width=80mm]{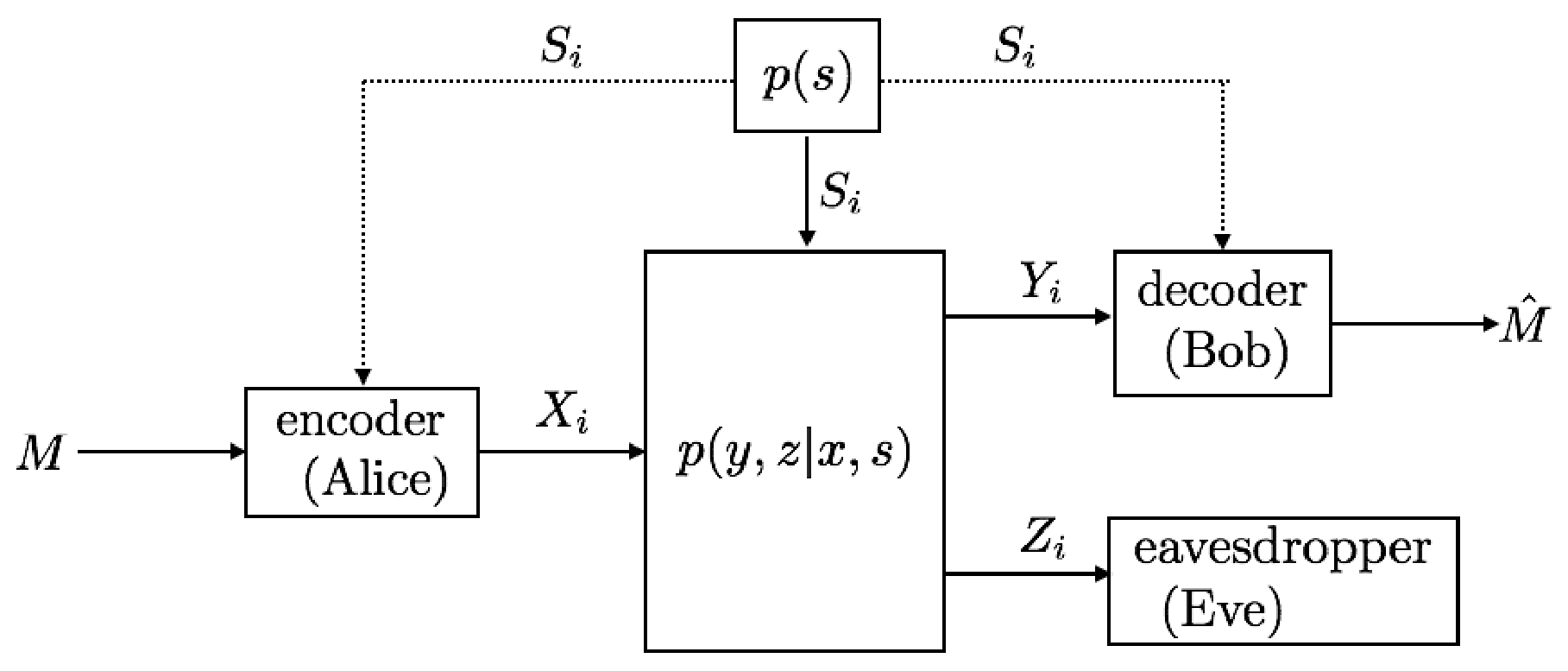}
\end{center}
\caption{WC with the same CSI available at Alice and Bob ($i=1,2,\cdots, n$).
          }
\label{fig-c3}
\end{figure}

\bchui\label{chui:both-enc1}
   %
%
%
Chia and El Gamal \cite{chia-elgamal} have considered the WC with common CSI available
  at both Alice and Bob as illustrated in Fig. \ref{fig-c3}. This channel, however, 
equivalently reduces  to 
that in Fig. \ref{fig-c2} with output $Y_S\equiv SY$ instead of $Y$.  
Then, since $H(S|UY_S)=H(S|USY)=0$,
$R_{\mbox{{\scriptsize\rm CSI-1}}}(p(u), p(x|u,s))$ and 
$R_{\mbox{{\scriptsize\rm CSI-2}}}(p(u), p(x|u,s))$ in (\ref{eq:rata-2}),  (\ref{eq:rata-3})
  reduce to
\beqn
R_{\mbox{{\scriptsize\rm CSI-1}}}(p(u), p(x|u,s))
& = & \min \Bigl[I(U; SY) -I(U; SZ) +H(S|Z), 
 I(U; SY)\Bigr],\label{eq:rata-12}\\
R_{\mbox{{\scriptsize\rm CSI-2}}}(p(u), p(x|u,s))
&=& \min \Bigl[H(S|UZ), I(U;SY)\Bigr]\label{eq:rata-13},
\eeqn
where the right-hand side of (\ref{eq:rata-12}) exactly coincides with the 
weak secrecy lower bound
\beqn
\min \Bigl[I(U; SY) -I(U; SZ) +H(S|Z), 
 I(U; SY)\Bigr]\label{eq:rata-12s-1}
 \eeqn
that was given by Chia and El Gamal \cite{chia-elgamal}, while 
the right-hand side of (\ref{eq:rata-13}) coincides with one more weak secrecy lower bound
 \beq
  \min \Bigl[H(S|UZ), I(U;SY)\Bigr]\label{eq:rata-12s-1OH}
  \eeq
  that was also given by \cite{chia-elgamal}.
Thus, Theorem \ref{teiri:main1} specialized to  the case with ``common" CSI available at both Alice 
and Bob provides
the {\em strong secrecy} version of their results.
Specifically, this concludes that Theorems 1, 2 and 3 in \cite{chia-elgamal} all hold with the strong secrecy criterion. \IEEEQED
\echui
\bchui\label{chui:typical}
A basic feature  of this paper is that  we do {\em not} invoke the argument of typical sequences
at all,
so we do not need the finiteness of alphabets $\cU, \cX, \cY, \cZ$, while 
the alphabet $\cS$ of  CSI $S$ needs to be finite.
\IEEEQED
\echui

\smallskip

\section{Proof of Theorem \ref{teiri:main1}}\label{intro-ge-cs1}

%
The whole coding scheme involves the 
transmission of 
 $b$ independent messages over the $b+1$ channel blocks  each of length $n$ 
 ($b$ is a sufficiently large  fixed positive integer),  which are indexed 
 by $j=0, 1,2,\cdots, b$.
The formal proof is provided in the sequel, where in block $j$ we let $\sU_j, \sS_j, \sX_j, \sY_j, \sZ_j$
(correlated i.i.d. sequences of length $n$ subject to joint probability $P_{USXYZ}$)
denote the random variables  to indicate
  channel input sequence, CSI sequence, channel input sequence for Alice,
channel output  sequences for Bob and Eve, respectively, 
whereas $M_j, M_{0j}, M_{1j}, $ $N_j$ denote the random variables to indicate
uniformly distributed confidential messages
to be sent, and auxiliary message, respectively.
Their realizations are indicated
by the corresponding lower case letters.

\bigskip

{\em Case A): Proof for the achievability of $R_{\mbox{{\scriptsize\rm CSI-1}}}$:}

\smallskip

\noindent
In what to follow, many kinds of (nonnegative) rates intervene with inequality constraints, 
which are listed as follows:
\beqn
\overline{R} & < & I(U; Y),\label{eq:deketa-1}\\
R &=& R_0 + R_1,\label{eq:deketa-2}\\
\overline{R}-R_0& > & I(U; SZ),\label{eq:deketa-3}\\
R_2 &>& H(S|UY),\label{eq:deketa-4}\\
R_0+R_1+R_2 &<& \overline{R},\label{eq:deketa-5}\\
R_1 +R_2 &<& H(S|Z).\label{eq:deketa-6f}
\eeqn
Fourier-Motzkin elimination (cf. El Gamal and Kim \cite{gamal-kim})
claims that the supremum of $R$ over all rates satisfying 
(\ref{eq:deketa-1})$\sim$ (\ref{eq:deketa-6f}) coincides with the right-hand side of (\ref{eq:rata-2}), so 
it suffices to show that rates $R$ satisfying 
(\ref{eq:deketa-1})$\sim$ (\ref{eq:deketa-6f})  are indeed achievable, where $\overline{R}$ is used to 
indicate an achievable rate for  usual channel coding (non-WC) between Alice and Bob.

\smallskip
 {\em Codebook generation:}
 
For each block $j\in [1:b]$,
split message $M_{j}\in [1:2^{nR}]$ into two independent {\em uniform} 
messages $M_{0j}\in [1:2^{nR_0}]$ and 
$M_{1j}\in [1:2^{nR_1}]$; thus $R=R_0+R_1$, 
where, in the process of channel transmission,  message 
$M_{0j}$ is protected by  WC coding,
and message $M_{1j}$ is protected by OTP cipher with the  secret key shared
using  CSI. The codebook generation consists of the following two parts:

\smallskip

 {\em 1) Message codebook generation:}

\begin{figure*}[!t]
\begin{center}
\includegraphics[width=120mm]{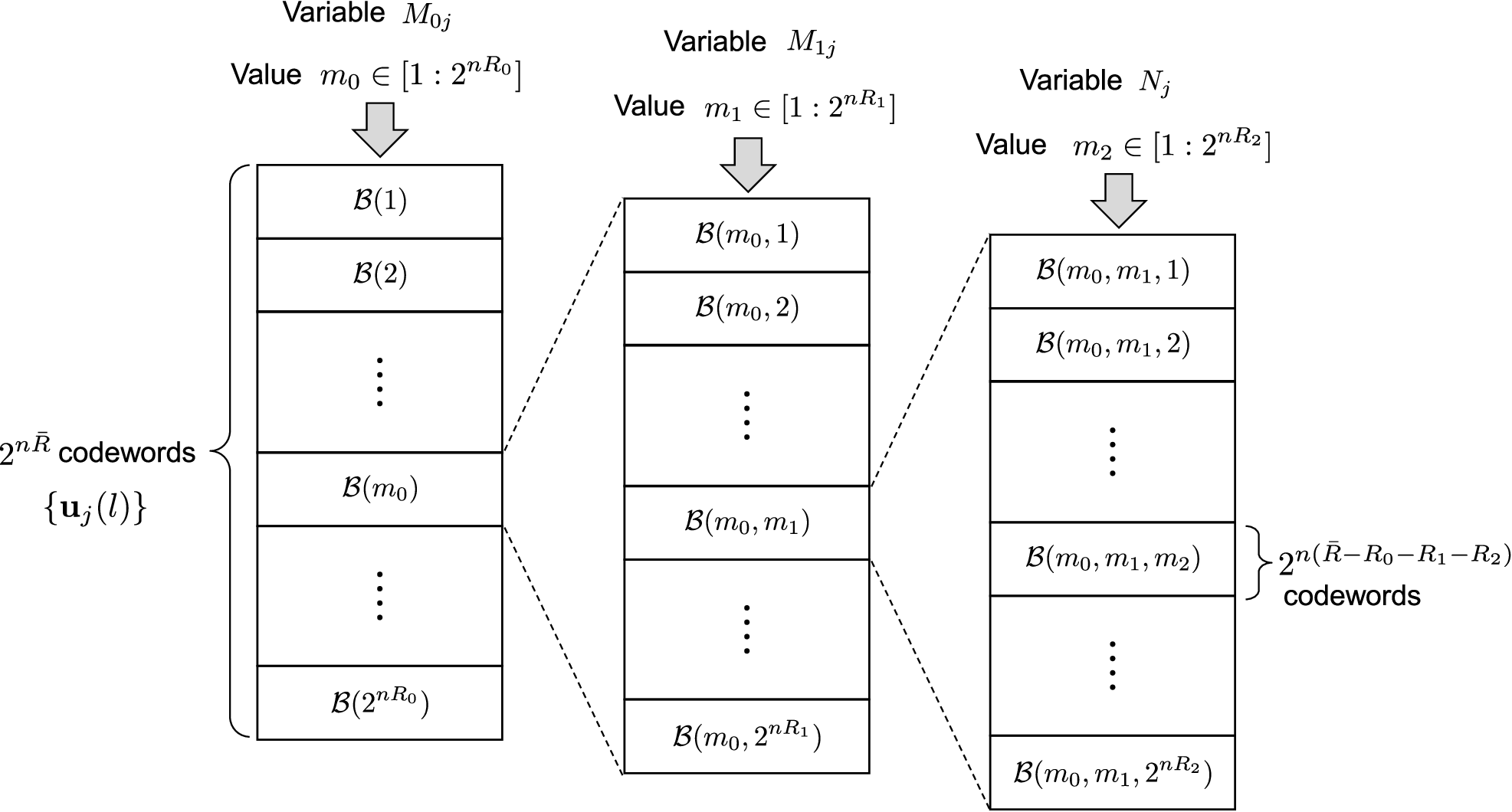}
\end{center}
\caption{Bin-partitioning for message codebook generation in each channel block $j$.
          }
\label{fig-c1-bin}
\end{figure*}

For each block $j\in [0:b]$, 
randomly and independently generate sequences $\ssu_j (l), l\in [1:2^{n\overline{R}}],$  each
according to  probability distribution $\prod_{i=1}^n p_U(u_i)\ (\ssu_j(l)=u_1u_2\cdots u_n)$. 
This is a random code and is denoted by $\cH_j$.
On the other hand, partition
the set $[1:2^{n\overline{R}}]$ of indices  into $2^{nR_0}$ equal-size bins $\cB(m_0), 
m_0\in [1: 2^{nR_0}]$. Moreover, partition the indices within each bin $\cB(m_0)$ into $2^{nR_1}$
equal-size sub-bins $\cB(m_0, m_1), m_1\in [1: 2^{nR_1}]$. 
Furthermore, partition the indices within each bin $\cB(m_0, m_1)$ into $2^{nR_2}$ 
equal-size sub-sub-bins $\cB(m_0, m_1, m_2), m_2\in [1: 2^{nR_2}]$ (cf. Fig. \ref{fig-c1-bin}).
These bins are all non-empty because of (\ref{eq:deketa-5}).

\smallskip

 {\em 2) Key codebook generation:}

In order to construct an efficient key $K_j=\kappa (\sS_j)$ of rate $R_1$ 
using the CSI $\sS_j$, we invoke the following two celebrated lemmas:
\bhodai[Slepian and Wolf \cite{slepian-wolf}]\label{hodai:deketa-1}
Let $\varepsilon >0$ be an arbitrarily small number and let $R_2 > H(S|UY)$ (cf. (\ref{eq:deketa-4})).
Then, there exists  ({\em deterministic}) functions 
$\sigma: \cS^n\to [1:2^{nR_2}]$ and 
$\phi:  [1:2^{nR_2}] \times \cU^n\times\cY^n\to \cS^n$ such that
\beq\label{eq:deketa-8s}
\Pr\{ \sS_j \neq \tilde{\sS}_j\}\le \varepsilon
\eeq
for all sufficiently large $n$, where $\tilde{\sS}_j=\phi(\sigma(\sS_j), \sU_j, \sY_j)$. 
 \IEEEQED 
\ehodai

For simplicity, we use also the notation $N_{j+1}\equiv \sigma (\sS_j)$, which is  the random variable 
conveying  the  auxiliary message used for generating the  common secret key between Alice and Bob.
\bhodai[\mbox{Csisz\'ar and K\"orner \cite[Corollary 17.5]{csis-kor-2nd}}]\label{hodai:deketa-2}
Let $\varepsilon >0$ be an arbitrarily small number and let 
$R_1+R_2 < H(S|Z)$ (cf. (\ref{eq:deketa-6f})).
Then, with the same $N_{j+1}\equiv \sigma (\sS_j)$ as in Lemma \ref{hodai:deketa-1},
 there exists a ({\em deterministic}) key function
$\kappa: \cS^n\to [1:2^{nR_1}]$ such that
\beq\label{eq:deketa-9}
{\sf S}\left(\kappa (\sS_j)\sigma(\sS_j)|\sZ_j\right) \le \vep
\eeq
for all sufficiently large $n$, where  we use the notation (called the {\em security index}):
\footnote{
Specifically, in the proof of this lemma, it suffices to  make  uniform random hashing
 $(\kappa, \sigma): \cS^n\to [1:2^{nR_1}]\times  [1:2^{nR_2}]$ 
(and hence uniform random binnning  $\sigma: \cS^n\to [1:2^{nR_2}]$ simultaneously) to construct
a pair of 
{\em deterministic} mappings $\kappa (\sS_j)\sigma(\sS_j) \equiv (\kappa (\sS_j),\sigma(\sS_j) )$ satisfying (\ref{eq:deketa-8s})  and (\ref{eq:deketa-9}).
This is possible owing to  rate constraints $R_2>H(S|UY)$ and $R_1 +R_2 <H(S|Z)$. 
}
\beq\label{eq:deketa-10}
{\sf S}(K|F)\stackrel{\Delta}{=}D(P_{KF}||Q_{K}\times P_F)
\eeq
with the uniform distribution  $Q_{K}$  on the range of $K$,
the KL divergence $D(\cdot||\cdot)$
and the product distribution $Q_{K}\times P_F$.
%
\IEEEQED 
\ehodai

We use the thus defined {\em deterministic} function $K_{j-1}\equiv\kappa(\sS_{j-1})$ as the key
 to be used  in the next block $j$. 

\smallskip

{\em Encoding scheme}

%
           %
            \begin{figure*}[!t]
\begin{center}
\includegraphics[width=120mm]{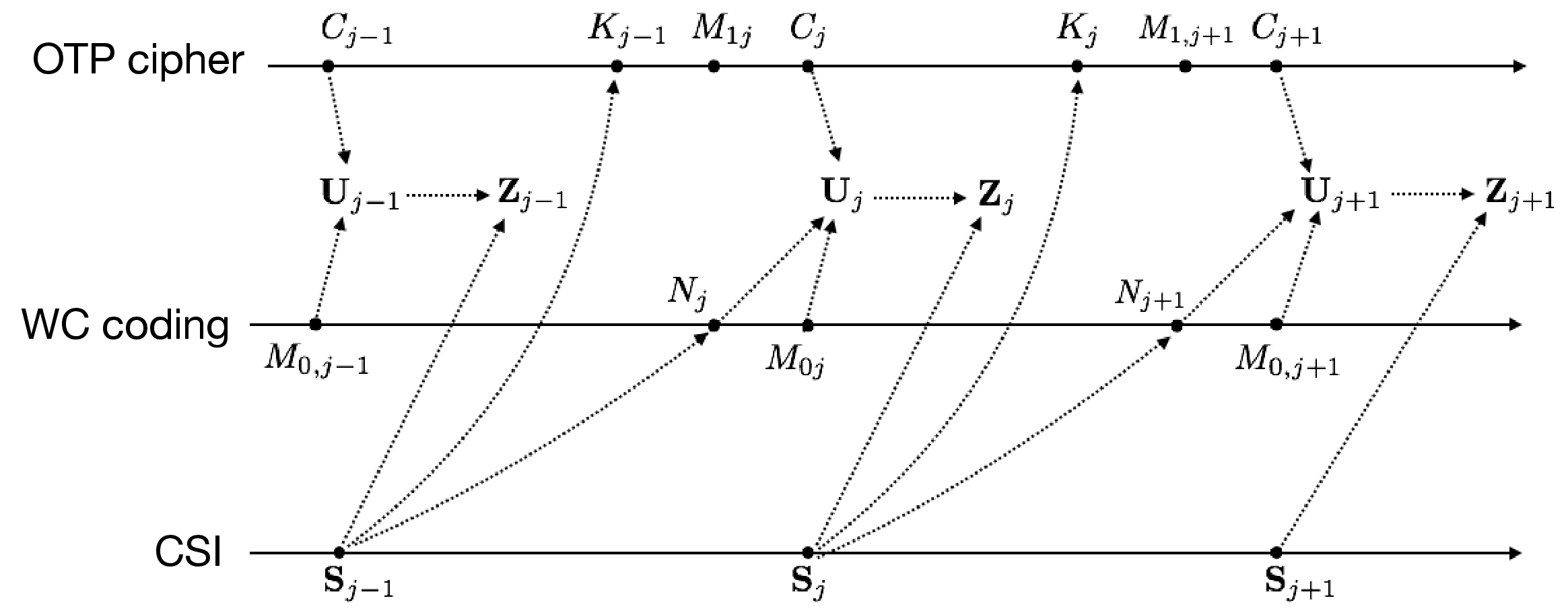}
\end{center}
\caption{Sequence diagram of block Markov coding  ($C_j=K_{j-1}\oplus M_{1j}$; $j=1,2, \cdots, b$).
          }
\label{fig-c1}
\end{figure*}

%


We  use the  block coding scheme as in Fig. \ref{fig-c1}, which is based on  the block Markov coding scheme 
invented by Cover and El Gamal \cite{cover-elgamal} (cf. Fig. \ref{fig-c1}) 
and applied to the WC with CSI by
Chia and El Gamal \cite{chia-elgamal}.
%
The first  block $j=0$ provides only the CSI sequence  
$\sS_0$ for Alice 
to
be used for encoding in the second block $j=1$  with $M_0=N_0=``1"$ (fixed dummy message).
%
In each block $j\in [1:b]$, given a message triple $(M_{0j}=m_0, 
M_{1j}=m_1, N_j=m_2)$,  Alice first computes $c_j=k_{j-1}\oplus m_1$ (mod $2^{nR_1}$) and let 
$L\stackrel{\Delta}{=} L(m_0, c_j, m_2)$ be 
the random index uniformly distributed on the bin $\cB(m_0, c_j, m_2)$ with 
 $k_{j-1}=\kappa(\sss_{j-1})$ as specified  in Lemma \ref{hodai:deketa-2}.
Alice then sends out for channel transmission a randomly generated sequence $\sX_j$ according to 
conditional probability 
$\prod_{i=1}^np_{X|US}(x_i|u_i(L), s_i)$, where 
$\ssx_j =x_1x_2\cdots x_n,$ $ \ssu_j (L) =u_1(L)u_2(L)\cdots u_n(L),$ $
\sss_j =s_1s_2\cdots s_n$.
We set $\sU_j=\ssu_j(L)$.

\smallskip
 {\em Decoding scheme and evaluation of probability of error:}
 
Let $\sY_j$ be the output for Bob due to $\sU_j$. Consider the stationary memoryless channel $\omega_n(\ssy|\ssu)\equiv P_{\sY_j|\sU_j}(\ssy|\ssu)$ with input $\ssu$ and output $\ssy$.
 For this channel we use the maximum likelihood decoding, that is,  we let $\hat{l}$ denote an index
 such that
\beq\label{eq:decode-1}
\omega_n(\ssy|\ssu_j(\hat{l}))= \max_{l\in [1:2^{n\overline{R}}]}\omega_n(\ssy|\ssu_j(l)),
\eeq
and set $\hat{\sU}_j=\ssu_j(\hat{l}).$ 
 Find the $(\hat{m}_0,  \hat{c}, \hat{m}_2)$ such that $\hat{l} \in \cB(\hat{m}_0,  \hat{c}, \hat{m}_2)$.
Next, compute $\hat{m}_{1}= \hat{c}\ominus  \hat{k}_{j-1}$ (mod $2^{nR_1}$),
where  $\hat{k}_{j-1}= \kappa(\hat{\sss}_{j-1})$
with $\hat{\sss}_{j-1}=\phi(\hat{m}_2,\ssu_{j-1}(\hat{l}), \ssy_{j-1})$ and we notice that
$\hat{\sss}_{j-1}=\tilde{\sss}_{j-1}$ if $\hat{m}_2=m_2$ and $\ssu_{j-1}(\hat{l}) =\ssu_{j-1}(L)$ 
 (cf. Lemmas \ref{hodai:deketa-1} and \ref{hodai:deketa-2}).
Finally, declare that the message pair $(\hat{m}_0,  \hat{m}_1)$ was sent.
In order to evaluate the  probability of decoding error 
\beq\label{eq:decode-1error}
Pe(j) \equiv \Pr\{(M_{0j}, M_{1j})\neq (\hat{M}_{0j}, \hat{M}_{1j})|\cH\},
\eeq
we invoke%
\bhodai[\mbox{Gallager \cite[Theorem 5.6.2]{gall}}]\label{hodai:deketa-3}
Let $\varepsilon >0$ be an arbitrarily small number and let $\overline{R}<I(U;Y)$ (cf. (\ref{eq:deketa-1})).
Then, 
\beq\label{eq:decode-1error-1}
\Pr\{(M_{0j}, C_j, N_j\equiv M_{2j}, \sU_j)\neq (\hat{M}_{0j},  \hat{C}_j, \hat{ N}_{j}\equiv \hat{M}_{2j}, 
\hat{\sU}_j)|\cH\} \le \vep
\eeq
for all sufficiently large $n$. \IEEEQED 
\ehodai

Then, in view of Lemmas \ref{hodai:deketa-1} and \ref{hodai:deketa-3}, we have
\beqn\label{eq:decode-1error-2}
Pe(j) &\le & \Pr\{(M_{0j}, C_j, N_j,\sU_j)\neq (\hat{M}_{0j},  \hat{C}_j, \hat{N}_{j}, \hat{\sU}_j)|\cH\}\nonumber\\
& &+ \Pr\{ \sS_{j} \neq \tilde{\sS}_{j}\}\nonumber\\
&\le &  2\vep.
\eeqn
Thus, it is concluded that the total probability of decoding error over all the $b$ blocks is less than or equal to $2b\vep$. It should be remarked here that the total transmission rate averaged over all $b+1$ blocks
is $\frac{bR}{b+1}$ because only the $b$ blocks of them are effective for message transmission, 
which can be made as close  to $R$ as desired by letting $b$ large enough. 
\smallskip

{\em Evaluation of information leakage:}

\smallskip

We use the following notation: for $j\in [1:b]$,
\beqns
\cH&=& \cH_1\cH_2\cdots \cH_b, \\
M_j&=& M_{0j}M_{1j},\\
M^j&=& M_1M_2\cdots M_j,\\
M^{[j]} &=& M_{j}M_{j+1}\cdots M_{b},\\
\sZ^j&=& \sZ_1\sZ_2\cdots \sZ_j,\\
\sZ^{[j]}&=& \sZ_j\sZ_{j+1}\cdots \sZ_b,\\
\eeqns
where we notice that $\sZ_j$ is the channel output for Eve in block $j$.
\bchui\label{chui:han1}
Since $K_{j-1}$ and $M_{1j}$ are independent and 
$M_{1j}$ is assumed to be uniformly distributed,
the OTP cipher  claims that $K_{j-1}$ and $C_j=K_{j-1}\oplus M_{1j}$
 are independent and  $C_j$ is uniformly distributed (cf. Shannon \cite{shannon-secrecy}).
 Notice here that $K_{j-1}$ is not necessarily uniformly distributed, and hence $M_{1j}$ and $C_j$
  are {\em not} necessarily independent.
On the other hand, $\sZ_{j-1}$ may affect $\sZ_j$ only through $K_{j-1}N_j$ 
and inversely  $\sZ_{j-1}$ may be affected by  $\sZ_j$ only through $C_jN_j$. 
 This property plays the crucial role in evaluating the performance of
 our coding scheme (cf. Fig.\ref{fig-c1}).
  \IEEEQED
\echui
%
%
%
%
\smallskip

In the sequel we show that the information leakage to Eve $I_E=I(M^b; \sZ^b|\cH)$ 
over the whole $b+1$ blocks
goes to zero as $n\to \infty$. 

To do so, we begin with
\beqn\label{eq:svbta1}
A &\stackrel{\Delta}{=} & I(M^b; \sZ^b |\cH)\nonumber\\
&=&\sum_{j=1}^b I(M_j; \sZ^b|M^{[j+1]}\cH) \nonumber\\
& \stackrel{(a)}{\le} & \sum_{j=1}^bI(M_j; \sZ^b|\sS_jM^{[j+1]}\cH)\nonumber\\
& \stackrel{(b)}{=} & \sum_{j=1}^bI(M_j; \sZ^j|\sS_j\cH)\nonumber\\
&=& \sum_{j=1}^bI(M_{0j}M_{1j}; \sZ^j|\sS_j\cH)\nonumber\\
&=& \sum_{j=1}^bI(M_{0j}M_{1j}; \sZ^{j-1}|\sS_j\cH)\nonumber\\
& & + \sum_{j=1}^bI(M_{0j}M_{1j}; \sZ_j|\sZ^{j-1}\sS_j\cH)\nonumber\\
& \stackrel{(c)}{=} &  \sum_{j=1}^bI(M_{0j}M_{1j}; \sZ_j|\sZ^{j-1}\sS_j\cH)\non\\
 &= & B + C
\eeqn
with
\beqn
B&\stackrel{\Delta}{=} &  \sum_{j=1}^bI(M_{0j}; \sZ_j|\sZ^{j-1}\sS_j\cH)\label{eq:svbta21}\\
C&\stackrel{\Delta}{=} &  \sum_{j=1}^bI(M_{1j}; \sZ_j|M_{0j}\sZ^{j-1}\sS_j\cH),\label{eq:svbta22}
\eeqn
where $(a)$ follows from the independence of $M_j$ and $\sS_j$ given $M^{[j+1]}\cH$; 
$(b)$ follows from 
the Markov chain property $M_j\to \sZ^j\sS_j \to \sZ^{[j+1]}M^{[j+1]}$ given $\cH$;
$(c)$ follows from the independence of $M_{0j}M_{1j}\sS_j$ and $\sZ^{j-1}$ given $\cH$.

Let us now separately evaluate $B$ and $C$ in (\ref{eq:svbta21}) and (\ref{eq:svbta22}). First, %
\beqn\label{isitaa-1}
B&=& \sum_{j=1}^bI(M_{0j}; \sZ_j|\sZ^{j-1}\sS_j\cH)\non\\
&\le &  \sum_{j=1}^bI(\sZ^{j-1}M_{0j}; \sZ_j|\sS_j\cH)\non\\
&=&   \sum_{j=1}^bI(M_{0j}; \sZ_j|\sS_j\cH)\non\\
& & +   \sum_{j=1}^bI(\sZ^{j-1}; \sZ_j|M_{0j}\sS_j\cH)\non\\
&\le &  \sum_{j=1}^bI(M_{0j}; \sZ_j|\sS_j\cH)\non\\
& & +   \sum_{j=1}^bI(\sZ^{j-1}; N_jM_{0j}\sS_j\sZ_j|\cH)\non\\
&\stackrel{(d)}{=}  &  \sum_{j=1}^bI(M_{0j}; \sS_j\sZ_j|\cH)\non\\
& & +   \sum_{j=1}^bI(N_j; \sZ^{j-1}|\cH),
\eeqn
where $(d)$ follows from the independence of  $M_{0j}$ and  $\sS_j$
 and from
the Markov chain property
$\sZ^{j-1}\to N_j \to M_{0j}\sS_j\sZ_j$ given $\cH$.

 Next, $C$ can be upper bounded as
\beqn\label{eq:svbta3a}
C&=& \sum_{j=1}^bI(M_{1j}; \sZ_j|M_{0j}\sZ^{j-1}\sS_j\cH)\non\\
&\le& \sum_{j=1}^bI(\sZ^{j-1}M_{1j}; \sZ_j|M_{0j}\sS_j\cH)\non\\
%
%
%
&=& D +E,
\eeqn
where 
\beqn
D&\stackrel{\Delta}{=} &  \sum_{j=1}^bI(M_{1j}; \sZ_j|M_{0j}\sS_j\cH)\label{eq:svbta4s}\\
E&\stackrel{\Delta}{=} &  \sum_{j=1}^bI(\sZ^{j-1}; \sZ_j|M_{0j}M_{1j}\sS_j\cH).
 \label{eq:svbta49}
\eeqn
Then,
\beqn
D &\le & \sum_{j=1}^bI(M_{1j}; C_j\sZ_j|M_{0j}\sS_j\cH)\non\\
 & =& F + G,\label{eq:mata-0}
 \eeqn
 where 
 \beqn
 F&=& \sum_{j=1}^bI(M_{1j}; C_j|M_{0j}\sS_j\cH),\non\\
 G&=& \sum_{j=1}^bI(M_{1j}; \sZ_j|M_{0j}\sS_jC_j\cH).\label{eq:mata-1}
 \eeqn
 Then, 
 \beqn
F &= &  \sum_{j=1}^bI(M_{1j}; C_j|M_{0j}\sS_j\cH)\non\\
&\stackrel{(f)}{=} &\sum_{j=1}^b I(M_{1j}; C_j)\non\\
&=& H(C_j) - H(C_j|M_{1j})\non\\
&\stackrel{(k)}{=}& H(C_j) - H(K_{j-1}|M_{1j})\non\\
& \stackrel{(g)}{=}&H(C_j)-H(K_{j-1})\non\\
&\stackrel{(p)}{=}& D(P_{K_{j-1}}||Q_{K_{j-1}}),
\label{eq:svbta4s11}
\eeqn
where
%
$(f)$ follows from the independence of  $M_{1j} C_j$ and $M_{0j}\sS_j\cH$;
$(k)$ follows from $K_{j-1}\oplus M_{1j} = C_j$;
$(g)$ follows from the independence of $K_{j-1}$ and $M_{1j}$;
$(p)$ follows from that $C_j$ is uniformly distributed on the range of $K_{j-1}$.
%
%

Moreover,
\beqn
G&=& \sum_{j=1}^bI(M_{1j}; \sZ_j|M_{0j}\sS_jC_j\cH)\non\\
&\stackrel{(e)}{=}& \sum_{j=1}^bI(K_{j-1} ;\sZ_j|M_{0j}\sS_jC_j\cH)\non\\
&\stackrel{(j)}{\le}& \sum_{j=1}^bI(K_{j-1} ;N_j|M_{0j}\sS_jC_j\cH)\non\\
&\stackrel{(m)}{=}& \sum_{j=1}^bI(K_{j-1} ;N_j),\label{eq:svb-abc}
\eeqn
where $(e)$ follows from $C_j=K_{j-1}\oplus M_{1j}$; $(j)$ follows from 
the data processing lemma using the Markov chain property
$K_{j-1}\to N_j\to \sZ_j$ given $M_{0j}\sS_jC_j\cH$;
$(m)$ follows from the independence of $K_{j-1}N_j$ and $M_{0j}\sS_jC_j\cH$.
On the other hand, 
\beqn\label{eq:svbta-op}
E&\le&  \sum_{j=1}^bI(\sZ^{j-1}; K_{j-1}N_j\sZ_j|M_{0j}M_{1j}\sS_j\cH)\non\\
& \stackrel{(h)}{=}&  \sum_{j=1}^bI(\sZ^{j-1}; K_{j-1}N_j|M_{0j}M_{1j}\sS_j\cH)\non\\
& \stackrel{(i)}{=}&  \sum_{j=1}^bI(K_{j-1}N_j; \sZ^{j-1}|\cH),
\eeqn
where $(h)$ follows from the Markov chain property $\sZ^{j-1}\to K_{j-1}N_j\to \sZ_j$ given $M_{0j}M_{1j}\sS_j\cH$;
$(i)$ follows from the independence of $ \sZ^{j-1}K_{j-1}N_j\cH$ and $M_{0j}M_{1j}\sS_j$.
\bigskip

%
%
\noindent
Thus, summarizing up (\ref{eq:svbta1})$\sim$ (\ref{eq:svbta-op}), we have
the upper bound on the information leakage to Eve $I_E=I(M^b;\sZ^b|\cH)$ as 
\bhodai[Information leakage bound]\label{hodai:hirai-1}
\beqn
I(M^b;\sZ^b|\cH) &\le& \sum_{j=1}^bI(M_{0j}; \sS_j\sZ_j|\cH) \label{eq:marchan-1}\\
& & + \sum_{j=1}^{b}I(N_j; \sZ^{j-1}|\cH).\label{eq:marchan-3-p}\\
& & + \sum_{j=1}^{b} I(K_{j-1}; N_j)\label{eq:marchan-002}\\
& &+ \sum_{j=1}^{b} D(P_{K_{j-1}}||Q_{K_{j-1}})\label{eq:marchan-2}\\
& & + \sum_{j=1}^{b}I(K_{j-1}N_j; \sZ^{j-1}|\cH).\label{eq:marchan-3}
\eeqn
\IEEEQED
\ehodai
Here, the first term $I(M_{0j}; \sS_j\sZ_j|\cH)$ specifies
the resolvability performance for Eve; the second term 
 $I(N_j; \sZ^{j-1}|\cH)$ specifies the inter-block interaction effect in the block Markov coding scheme; 
 the third and fourth terms 
$ I(K_{j-1}; N_j)$, $D(P_{K_{j-1}}||Q_{K_{j-1}})$
specify
 the key performance for Bob; and the fifth term $I(K_{j-1}N_j; \sZ^{j-1}|\cH)$ specifies 
 the key performance for Eve.

The third and fourth ones  are evaluated as follows. 
We can rewrite the security index
${\sf S}(\kappa(\sS_j)\sigma (sS_j)|\sZ_j)$ in (\ref{eq:deketa-9}) of Lemma \ref{hodai:deketa-2} as
\beqn\label{eq:deketa-9sk}
\lefteqn{{\sf S}(\kappa(\sS_j)\sigma (\sS_j)|\sZ_j)}\non\\
&\ge & D(P_{\kappa (\sS_j)\sigma (\sS_j)}||Q_{\kappa (\sS_j)}\times Q_{\sigma (\sS_j)})\non\\
 &=&D(P_{\kappa (\sS_j)\sigma (\sS_j)}||P_{\kappa (\sS_j)}\times P_{\sigma (\sS_j)})\non\\
& & + D(P_{\kappa (\sS_j)}||Q_{\kappa (\sS_j)}) + 
    D(P_{\sigma (\sS_j)}||Q_{\sigma (\sS_j)})  \non\\
 %
  & \ge &D(P_{\kappa (\sS_j)\sigma (\sS_j)}||P_{\kappa (\sS_j)}\times P_{\sigma (\sS_j)})\non\\
  & = &I(\kappa (\sS_j); \sigma (\sS_j))\non\\
   &=&I(K_j; N_{j+1})\label{eq:deketa-9sk-1}.
\eeqn
%
%
Moreover, 
\beqn\label{eq:deketa-9s}
\lefteqn{{\sf S}(\kappa(\sS_j)\sigma (\sS_j)|\sZ_j)}\non\\
 &\ge& {\sf S}(\kappa(\sS_j)|\sZ_j)\non\\
 & = &D(P_{K_{j}}||Q_{K_{j}})+I(K_j; \sZ_j)\non\\
  & \ge &D(P_{K_{j}}||Q_{K_{j}}).
\eeqn
Therefore, Lemma \ref{hodai:deketa-2} claims that
\beqn
I(K_{j-1}; N_{j}) &\le& \vep, \label{eq:svbta114q}\\
D(P_{K_{j-1}}||Q_{K_{j-1}})&\le &\vep.\label{eq:svbta114q-2}
\eeqn

In order to evaluate the second and fifth ones, we use the following lemma, which is the Alice-only CSI counterpart of 
\cite[Proposition 1]{chia-elgamal}:
\bhodai[Key secrecy lemma]\label{hodai:marchan}
Let $\varepsilon >0$ be an arbitrarily small number and let $R_1+R_2 <H(S|Z)$
 (cf. (\ref{eq:deketa-6f})).
Then, for $j\in [1: b]$,
\beqn
& \mbox{i)} & I(K_{j-1}N_{j}; \sZ_{j-1}|\cH)  \le \vep,\label{hodai-ma2}\\
& \mbox{ii)} & I(K_{j-1}N_{j}; \sZ^{j-1}|\cH) \le b\vep \label{hodai-ma2va}
\eeqn
for all sufficiently large $n$. \IEEEQED 
\ehodai
{\em Proof}: See Appendix \ref{addA}.\\
From  (\ref{hodai-ma2va}) we immediately have
\beqn\label{eq:machan-7}
     I(N_{j}; \sZ^{j-1}|\cH)   \le I(K_{j-1}N_{j}; \sZ^{j-1}|\cH) \le b\vep.
\eeqn

Now, what remains to be done is to evaluate the first one $I(M_{0j}; \sS_j\sZ_j|\cH)$. To do so, we invoke
the following resolvability lemma:
\bhodai[\mbox{Resolvability lemma}]\label{hodai:deketa-5}
Let $\varepsilon >0$ be an arbitrarily small number and let $\overline{R}-R_0>$ $
I(U;SZ)$ (cf. (\ref{eq:deketa-3})).
Then, 
\beq\label{eq:decode-1error-2ws}
I(M_{0j}; \sS_j\sZ_j|\cH) \le \vep
\eeq
for all sufficiently large $n$. \IEEEQED 
\ehodai

{\em Proof:} See Appendix \ref{addB}. \IEEEQED

\smallskip
An immediate consequence of Lemma \ref{hodai:hirai-1} together with (\ref{eq:svbta114q}), 
(\ref{eq:svbta114q-2}),
(\ref{eq:machan-7}) and
 (\ref{eq:decode-1error-2ws}) is
\beq\label{eq:decode-1error-1g2}
I(M^b;\sZ^b|\cH) \le (3b+2b^2) \vep,
\eeq
thereby completing the proof for {\em Case A)}. \IEEEQED

\bigskip 

\noindent
{\em Case B): Proof for the achievability of $R_{\mbox{{\scriptsize\rm CSI-2}}}$:}

%
%

%
%
%


The remainder of Theorem \ref{teiri:main1} to be proved is the acievability
of $R_{\mbox{{\scriptsize\rm CSI-2}}}(p(u), p(x|u,s))$ in (\ref{eq:rata-3}).

The rate constraints in this case  are listed as follows ($R_0=0$):
\beqn
\overline{R} & < & I(U; Y),\label{eq:deketa-1s}\\
R &=&  R_1,\label{eq:deketa-2s}\\
R_2 &>& H(S|UY),\label{eq:deketa-4s}\\
R_1+R_2 &<& \overline{R},\label{eq:deketa-5s}\\
R_1+R_2 &<&H(S|UZ).\label{eq:deketa-5sf}
\eeqn
%
These constraints are the same as those in {\em Case A)} with $R_0=0$
and $H(S|UZ)$ instead of $H(S|Z)$,
 where 
constraint (\ref{eq:deketa-3}) is not necessary here because of $R_0=0$.
The reason for the replacement of $H(S|Z)$ by $H(S|UZ)$ is that, since $R_0=0$, 
we cannot here leverage the randomization (over input $\sU_j$) due to 
 Wyner's WC coding  to keep the  $\sU_j$ secure  from the attack by Eve.
Fourier-Motzkin elimination 
claims that the supremum of $R$ over all rates satisfying 
(\ref{eq:deketa-1s})$\sim$ (\ref{eq:deketa-5sf}) coincides with the 
$R_{\mbox{{\scriptsize\rm CSI-2}}}(p(u), p(x|u,s))$, 
so 
it suffices to show that rates $R$ satisfying 
(\ref{eq:deketa-1s})$\sim$ (\ref{eq:deketa-5sf})  are  achievable.

 In this case too, the  proof argument  parallels  those as  developed in the proof for {\em Case A)} 
with $R_0=0$, where we notice that
  $I(M_{0j}; \sS_j\sZ_j|\cH)=0$ in 
Lemma \ref{hodai:hirai-1} and hence Lemma \ref{hodai:deketa-5} is not needed here, thereby
completing the achievability proof for this case. \IEEEQED

%
%
%
\section{Secrecy capacity results}\label{sec:illust}
Thus far we have developed  achievability arguments for WCs with CSI available only at the encoder (Alice)
to establish Theorem \ref{teiri:main1} on lower bounds to 
the  secrecy capacity $C_{\mbox{{\scriptsize\rm CSI-E}}}$.
In this section, in order to get more  insights  into this theorem, we 
address the problem of bounding the  secrecy capacities for the case of ({\em statistically}) degraded  WCs,
which is an important class of WCs.

%

\smallskip
 
 \  \ Let us first describe the first theorem in this section:

 \bteiri\label{teiri:RR-1}  For any degraded WC 
 ($Z$ is a degraded version of $Y$) with causal CSI only at Alice, we have
 \beqn
  C_{\mbox{{\scriptsize\rm CSI-E}}} 
 & \ge & \max_{p(x|s)}\min \bigl(I(XS;Y) - I(XS;Z), I(XS;Y) -H(S)\bigr)
 \label{eq:RR2}\\
C_{\mbox{{\scriptsize\rm CSI-E}}}
 & \le & \max_{p(x|s)}\min \bigl(I(XS;Y) - I(XS;Z), I(XS;Y) -I(S;Y)\bigr)
 \label{eq:RR1}\\
 C_{\mbox{{\scriptsize\rm NCSI-E}}}^{\mbox{{\scriptsize\sf K}}}
 & \ge& \max_{\scriptsize\scriptsize\scriptsize\scriptsize 
 I(XS;Y) \ge H(S)}\bigl(I(XS;Y) - I(XS;Z) \bigr)
 \label{eq:RR3}\\
 C_{\mbox{{\scriptsize\rm NCSI-E}}}^{\mbox{{\scriptsize\sf K}}}           
 & \le & \max_{p(x|s)} \bigl(I(XS;Y) - I(XS;Z) \bigr)
 \label{eq:RRM1},
 \eeqn
where $C_{\mbox{{\scriptsize\rm NCSI-E}}}^{\mbox{{\scriptsize\sf K}}}$ denotes 
 the non-causal {\em secret-key capacity} (as for the definition, see, e.g.,  Khisti {\em et al.} \cite{khisti-wornell}, 
 Prabhakaran {\em et al.} \cite{prabhakaran}, Bunin {\em et al.} \cite{bunin}).
 In contrast with this, $C_{\mbox{{\scriptsize\rm CSI-E}}}$
  may be called the {\em secret-message capacity}.
  The maximization in (\ref{eq:RR3}) is taken over all $XS$ such that $ I(XS;Y) \ge H(S)$.
  \IEEEQED
 \eteiri\label{chui:reiwa1}
 \bchui
{\rm Lower bounds (\ref{eq:RR2}) and (\ref{eq:RR3}) hold without the assumption of degradedness.
It is is easy to check that
 $I(XS;Y)-I(XS;Z)$ 
 in  (\ref{eq:RR2}) $\sim$  (\ref{eq:RRM1}) is nonnegative  for  degraded WCs,
while  $I(XS;Y)-H(S)$ in  (\ref{eq:RR2})  may be negative.
\IEEEQED
} 
 \echui

\smallskip
 {\em Proof of (\ref{eq:RR2}) (Achievability):}
 
Let $(X,S)$ be arbitrarily given, then the functional representation lemma \cite{gamal-kim} claims that
there exist a random variable $U$ and a deterministic function $f: \cU\times \cS\to \cX$
such that $U$ and $S$ are independent and $X=f(U, S)$. 

Then, 
the first term of the achievable rate $R_{\mbox{{\scriptsize\rm CSI-1}}}(p(u), p(x|u,s))$
 given  in Theorem \ref{teiri:main1}
can be rewritten as follows.
\beqn
\lefteqn{I(U; Y) -I(U; SZ)+ H(S|Z)-H(S|UY)}\nonumber\\
 &=&I(U; SY) -I(U;S|Y)-I(U; SZ)+ H(S|Z)-H(S|UY)\nonumber\\
  &=&I(U; Y|S) -I(U;Z|S)+ H(S|Z)-H(S|Y)\nonumber\\
    &\stackrel{(v)}{=}&I(XU; Y|S) -I(XU;Z|S)+ H(S|Z)-H(S|Y)\nonumber\\
    &\stackrel{(w)}{=}&I(X; Y|S) -I(X;Z|S)+ H(S|Z)-H(S|Y)\nonumber\\
    &=& I(XS;Y) - I(XS;Z),  \label{eq:RR4}
\eeqn
 where $(v)$ follows from that $X$ is  a deterministic function of $(U,  S)$; 
  $(w)$ follows from that $U\to SX\to YZ$ forms a Markov chain.

On the other hand, the second term of  $R_{\mbox{{\scriptsize\rm CSI-1}}}(p(u), p(x|u,s))$ 
can be rewritten as follows.
\beqn
\lefteqn{I(U; Y) -H(S|UY)}\nonumber\\
 &=&I(U; SY) -I(U;S|Y)-H(S|UY)\nonumber\\
  &=&I(U; Y|S) -H(S|Y)\nonumber\\
  &\stackrel{(y)}{=}& I(XU; Y|S) -H(S|Y)\nonumber\\
  &\stackrel{(z)}{=}&I(X;Y|S) -H(S|Y)\nonumber\\
  &=& I(XS;Y)-H(S),
     \label{eq:RR5}
\eeqn
 where in $ (y), (z)$ we have used the similar argument  to $ (v), (w)$.
 
Therefore, in view of Theorem  \ref{teiri:main1},
combining (\ref{eq:RR4}) and (\ref{eq:RR5})
 yields (\ref{eq:RR2}).

 {\em Proof of (\ref{eq:RR1}) (Converse):}

Here, we  invoke the following simple but powerful lemma:
    
    \bhodai[\mbox{Chen and Vinck \cite{chen-vinck}}]\label{hodai:state-enc-2}
    Let us consider a degraded WC with CSI $S$ such that $Z$  is a degraded version of $Y$. Then,
    the secrecy capacity with 
    {\em non-causal} CSI only at the encoder (=E), denoted by 
      $C_{\mbox{{\scriptsize\rm NCSI-E}}}$,   is upper bounded as
    \beq\label{eq:non-causal-conv-as2}
    C_{\mbox{{\scriptsize\rm NCSI-E}}} \le \max_{p(u|s)p(x|u,s)}(I(U; Y) - I(U; Z)),
    \eeq
    where  we notice that $U$ and $S$ may be  correlated.
    \IEEEQED
\ehodai

We compute $I(U;Y)$ and $I(U;Z)$ separately with arbitrary $USX$.
\beqn
\lefteqn{I(U; Y)}\nonumber\\
&=&I(USX; Y) -I(SX;Y|U)\nonumber\\
  &=&I(S;Y) +I(UX;Y|S)-I(S;Y|U)-I(X;Y|US)\nonumber\\
   &=&I(X; Y|S) +I(S;Y) -I(S;Y|U) - I(X;Y|US).  \label{eq:RR6}
\eeqn
Similarly,
\beqn
\lefteqn{I(U; Z)}\nonumber\\
   &=&I(X; Z|S) +I(S;Z) -I(S;Z|U) - I(X;Z|US).  \label{eq:RR6}
\eeqn
Hence, 
\beqn
\lefteqn{I(U;Y)-(U; Z)}\nonumber\\
   &=& 
   I(X;Y|S) - I(X;Z|S) + I(S;Y)-I(S;Z)\nonumber\\
   & & -\bigl(I(S;Y|U) - I(S;Z|U)\bigr) -\bigl(I(X;Y|US)-I(X;Z|US)\bigr)\nonumber\\
   &\le& I(X;Y|S) - I(X;Z|S) + I(S;Y)-I(S;Z)\nonumber\\
     &=& I(X;Y|S) - I(X;Z|S) +H(S|Z) - H(S|Y)\nonumber\\
      &=& I(XS;Y) - I(XS;Z),
      \label{eq:RR7}
\eeqn
where in the above inequality we have used the propertry $I(S;Y|U) - I(S;Z|U)\ge 0$ and $I(X;Y|US)-I(X;Z|US) \ge 0$,
which comes from the assumed degradedness. 

Another upper bound $R\le I(SX;Y)-I(S;Y)$ is derived as follows.
For any achievable rate $R$,  Fano inequality yields (with $\vep_n \to 0$ as $n$ tends to $\infty)$:
\beqn\label{eq:RR8}
nR & =& H(M)\nonumber\\
&\le &H(M) - H(M|Y^n) + n\vep_n\nonumber\\
&=& I(M;Y^n) + n\vep_n\nonumber\\
&=& \sum_{i=1}^nI(M; Y_i|Y^{i-1})+n\vep_n\nonumber\\
&\le& \sum_{i=1}^nI(MY^{i-1}; Y_i)+n\vep_n\nonumber\\
&\le& \sum_{i=1}^nI(MY^{i-1}; S_iY_i)+n\vep_n\nonumber\\
 &\stackrel{(p)}{=}& \sum_{i=1}^nI(MY^{i-1}; Y_i|S_i)+n\vep_n\nonumber\\
  &\le & \sum_{i=1}^nI(X_iMY^{i-1}; Y_i|S_i)+n\vep_n\nonumber\\
   &\stackrel{(q)}{=}& \sum_{i=1}^nI(X_i; Y_i|S_i)+n\vep_n\nonumber\\
     &\stackrel{(r)}{=}& nI(X_J; Y_J|S_JJ)+n\vep_n\nonumber\\
       &\le& nI(JX_J; Y_J|S_J)+n\vep_n\nonumber\\
        &\stackrel{(s)}{=}& nI(X_J; Y_J|S_J)+n\vep_n\nonumber\\
 &\stackrel{(t)}{=}& nI(X; Y|S)+n\vep_n,
\eeqn
where $(p)$ comes from the independence of $S_i$ and $MY^{i-1}$; $(q)$ follows from the Markov chain property
$MY^{i-1}\to X_iS_i\to Y_i$;  in $(r)$ $J$ is the random variable such that $\Pr\{J=i)\} = \frac{1}{n} \  (i=1,\cdots, n)$;
$(s)$ follows from the Markov chain property
$J\to X_JS_J\to Y_J$; in $(t)$ we have set $X=X_J, Y=Y_J, S=S_J$.

An immediate  consequence (deviding by $n$ and letting $n\to \infty$)  of (\ref{eq:RR8}) is 
\beqn\label{eq:RR9}
 R &\le& I(X; Y|S)\nonumber\\
 &=& I(XS;Y)-I(S;Y)
 \eeqn
 with  input\footnote{Actually, in order to conclude (\ref{eq:RR1}), we need to show that
 $X$ in (\ref{eq:RR7}) and $X$  (\ref{eq:RR9}) can be taken to be the same. However, this can be ascertained
 by carefully scritinizing the proof of Lemma \ref{hodai:state-enc-2}.} 
  $X$. Thus, combining (\ref{eq:RR7}) and (\ref{eq:RR9}) together with 
  Lemma \ref{hodai:state-enc-2} yields (\ref{eq:RR1}). 
 \IEEEQED

 {\em Secret-key capacity results:}

We see that there is a gap between the second terms of  (\ref{eq:RR2}) and  (\ref{eq:RR1}),  i.e.,   $H(S) \neq I(S;Y)$. 
 These terms are due to the physical channel capability limitation, which are indispensable when we are concerned 
 with the {\em secret-message capacity}
 like in the foregoing.  On the other hand, however, 
  as far as we are concerned with the {\em secret-key capacity}, such terms are not necessarily involved.
%

{\em Proof of  (\ref{eq:RR3}) (Achievability):}

 We first invoke the following achievability theorem:
\bteiri[Khisti {\em et al.} \cite{khisti-wornell}]\label{teiri:bunin}
For any WC,
the (weak) secret-key capacity
%
%
 with  {\em non-causal} CSI available only at the encoder 
 is lower bounded as 
\beq\label{eq:bubin-1}
C_{\mbox{{\scriptsize\rm NCSI-E}}}^{\mbox{{\scriptsize\sf K}}}\ge
 \max_{\scriptsize I(V;Y)\ge I(V;S)}
  (I(V;Y)-I(V;Z)),
\eeq
where  the maximization in (\ref{eq:bubin-1}) is taken over all $VS$ such that $I(V;Y)\ge I(V;S)$ and 
we notice   that $V$ and $S$ may be correlated. 
\IEEEQED
\eteiri
\bchui\label{chui:nebaneba-1}
In fact, the ``causal" version of formula (\ref{eq:bubin-1}) in Theorem \ref{teiri:bunin} is given by%
\beq\label{eq:bubin-13}
C_{\mbox{{\scriptsize\rm CSI-E}}}^{\mbox{{\scriptsize\sf K}}}\ge
 \max_{ I(V;Y)\ge I(V;S)}
  (I(V;Y)-I(V;Z)),
\eeq
where $V=(U, S)$ ($U$ and $S$ are independent) and 
$C_{\mbox{{\scriptsize\rm CSI-E}}}^{\mbox{{\scriptsize\sf K}}}$ denotes the (strong)
secret-key capacity with {\em causal} CSI available only at the encoder. 
Accordingly, $C_{\mbox{{\scriptsize\rm NCSI-E}}}^{\mbox{{\scriptsize\sf K}}}$ in (\ref{eq:RR3}) and 
 (\ref{eq:RRM1})  can be replaced by
$C_{\mbox{{\scriptsize\rm CSI-E}}}^{\mbox{{\scriptsize\sf K}}}$. 
The proof of (\ref{eq:bubin-13}) will be  given in a forthcoming paper
\cite{han-sasaki-revisited}
as a special case of 
 more general {\em causal}  WCs.
\IEEEQED
\echui

Now, let $(X,S)$ be arbitrarily given and let $U$ and $f$ be those as specified by  the functional representation lemma \cite{gamal-kim} as in the proof of (\ref{eq:RR2}).
%
%
We  then  compute the right-hand side of (\ref{eq:bubin-1}) with $V = (U,S) $ as follows:
\beqn\label{eq:bunin}
\lefteqn{I(US;Y)-I(US;Z)}\nonumber\\
&\stackrel{(b)}{=} & I(USX;Y) -I(USX;Z)\nonumber\\
&\stackrel{(c)}{=} & I(XS;Y) -I(XS;Z),
\eeqn
where in $(b)$ we noticed that $X$ is a deterministic function of $(U,S)$;
 $(c)$ follows from that $U\to XS \to YZ$ forms a Markov chain.  
 On the other hand, 
\beqn\label{eq:RRM2}
\lefteqn{I(US;Y)-I(US;S)}\nonumber\\
 &=& 
 I(SU;Y) -H(S)\nonumber\\
 &\stackrel{(d)}{=}& I(XSU;Y)-H(S)\nonumber\\
 &\stackrel{(e)}{=}& I(XS;Y)-H(S),  
\eeqn
 where $(d)$ follows since  $X$ is a deterministic function of $(U, S)$; $(e)$ 
 follows from the Markov chain property $U\to XS \to Y$. Thus,
 Theorem \ref{teiri:bunin} together with  (\ref{eq:bunin}) and (\ref{eq:RRM2})
 yields (\ref{eq:RR3}).
 %
 %

 \smallskip
 {\em Proof of (\ref{eq:RRM1}) (Converse):}

To show the converse part, we first observe that  Lemma \ref{hodai:state-enc-2} is still valid with
$C_{\mbox{{\scriptsize\rm NCSI-E}}}^{\mbox{{\scriptsize\sf K}}}$ instead of 
$C_{\mbox{{\scriptsize\rm NCSI-E}}}$,
which can be ascertained by carefully scrutinizing the proof in \cite{chen-vinck} 
(with secret key $K$ instead of secret message $M$) of Lemma \ref{hodai:state-enc-2}.
Then, in the entirely same way as  above, we have (\ref{eq:RR7}),
implying the converse here.
\IEEEQED

\smallskip
An immediate consequence of Theorem \ref{teiri:RR-1} is the following corollaries with degraded WCs,
where, hereafter,  we denote by $C_{\mbox{{\scriptsize\rm CSI-ED}}}, C_{\mbox{{\scriptsize\rm NCS-ED}}},
C^{\mbox{\scriptsize\sf K}}_{\mbox{{\scriptsize\rm NCSI-ED}}}$
the (strong) secrecy capacities of WCs with common CSI $S$ available at both the enceder (=E) and decoder  (=D):

\bkei[Strengthening of Chia and El Gamal \cite{chia-elgamal}]\label{kei:RR1} It holds that
\beqn\label{eq:RS1}
C_{\mbox{{\scriptsize\rm CSI-ED}}}
 &=&  C_{\mbox{{\scriptsize\rm NCSI-ED}}} \nonumber\\
 &=& \max_{p(x|s)}\min \bigl(I(XS;YS) - I(XS;Z), I(XS;YS) -H(S)\bigr)\nonumber\\
&=& \max_{p(x|s)}\min \bigl(I(X;Y|S) - I(X;Z|S)+H(S|Z), I(X;Y|S) \bigr)
 \eeqn
\ekei
 \bkei
 \label{kei:RR2}  It holds that
\beqn\label{eq:RS8}
   C^{\mbox{\scriptsize\sf K}}_{\mbox{{\scriptsize\rm CSI-ED}}} 
   &=& C^{\mbox{\scriptsize\sf K}}_{\mbox{{\scriptsize\rm NCSI-ED}}}\nonumber\\
   &=& \max_{p(x|s)} \bigl(I(XS;YS) - I(XS;Z) \bigr)\nonumber\\
&=& \max_{p(x|s)} \bigl(I(X;Y|S) - I(X;Z|S)+H(S|Z) \bigr).
 \eeqn
\ekei
 {\em \ \ Proof:} 
  It suffices to replace $Y$ by $SY$
   in  (\ref{eq:RR2})  $\sim$ (\ref{eq:RRM1}),
  where we have 
  taken account of Remark \ref{chui:nebaneba-1}.
 \IEEEQED.
 
 \bchui\label{chui:RS1}
 In fact, Khisti {\em et al.} \cite{khisti-wornell}
has, instead of (\ref{eq:RS8}), given the following (weak) formula ({\em not} assuming the degradedness) as:
\beqn\label{eq:RS2}
  C^{\mbox{\scriptsize\sf K}}_{\mbox{{\scriptsize\rm NCSI-ED}}}
= 
 \max_{p(u,x|s)} \bigl(I(U;Y|S) - I(U;Z|S)+H(S|Z) \bigr).
 \eeqn
However, the proof in \cite{khisti-wornell} for  the converse part   seems to contain a serious technical flaw.
\IEEEQED
\echui

\smallskip

%
%
%
 {\rm
 Next, following  Chia and El Gamal \cite{chia-elgamal}, let us consider the  following special WC to have
 \bkei\label{kei:han-sasaki}
  Let us consider a degraded WC such that $Z$ is a degraded version of $Y$ and
 $p(y, z|x, s) = p(y,z|x)$, then we have 
 \beq\label{yatto-q4}
C_{\mbox{{\scriptsize\rm CSI-E}}}  = C_{\mbox{{\scriptsize\rm NCSI-E}}} 
= \max_{p(x)}(I(X;Y)-I(X;Z)).
\eeq 
 \ekei
 \bchui\label{chui:kimp}
 {\rm 
  This result 
coincides with an intuition that this WC  may reduce simply to a  WC {\em without}
CSI at Alice and Bob, because CSI $S$ at Alice has no correlation to Bob.
In this connection, it will be useful to compare
 this result with that in \cite{chia-elgamal} with common CSI $S$ available at both the encoder
and decoder, the secrecy capacity
of which is given as 
\beq\label{yatto-q5}
C_{\mbox{{\scriptsize\rm CSI-ED}}}  = C_{\mbox{{\scriptsize\rm NCSI-ED}}} 
= \max_{p(x)}\min[I(X;Y)-I(X;Z)+H(S), I(X;Y)].
\eeq
 Clearly, in (\ref{yatto-q5}) 
 the state information $S$ 
 contributes to  making achievable rates higher
 by $H(S)$, whereas in (\ref{yatto-q4})  the CSI makes no contribution.
 This shows that ``two-sided" CSI  (available both at Alice and Bob)
 indeed can outperform 
 ``one-sided" CSI (available only at Alice).  
 }
 \IEEEQED
 \echui

 \smallskip
 \noindent
 {\it Proof of Corollary \ref{kei:han-sasaki}:}   
 We first observe that
  \beqn\label{eq:fuji-er-ya2}
 R_{\mbox{{\scriptsize CSI-0}}}(p(u), p(x|u,s)) &=& I(U;Y)-I(U;Z)\non\\
 &=&  I(X;Y)-I(X;Z)
 \eeqn
 by setting $X=U$ with $S$ independent of $X$, which implies the achievability.
 
In order to show the converse part, we compute 
 as follows:
 \beqn\label{eq:fuji-er-ya3}
 \lefteqn{I(U;Y)-I(U;Z)}\non\\
 &=& I(UX; Y)-I(X;Y|U)\non\\
 & & - I(UX; Z)+I(X;Z|U)\non\\
 &=& I(X;Y)-I(X;Z)\non\\
 & & -I(X; Y|U)+I(X;Z|U).
 \eeqn
On the other hand, owing to the assumed degradedness, we have
 \beqn\label{eq:fuji-er-ya4}
 \lefteqn{I(X;Y|U)}\non\\
 &=& I(X; ZY|U)\non\\
 &=& I(X;Z|U)+I(X;Y|UZ).
 \eeqn
From  (\ref{eq:fuji-er-ya3}) and (\ref{eq:fuji-er-ya4}), it follows that
\beq\label{yatto-q2}
I(U;Y)-I(U;Z) \le I(X;Y)-I(X;Z).
\eeq
Thus, in light of Lemma \ref{hodai:state-enc-2}  together with (\ref{eq:fuji-er-ya2}) and (\ref{yatto-q2}),
the corollary is concluded.  \IEEEQED

 %

 \smallskip

So far, we have studied WCs with {\em non-binary} alphabets. It would also be interesting
to see what happens with {\em binary} WCs ($\cU=\cX=\cY=\cZ=\cS=\{0,1\}$).
  {\rm
     %
   Letting $\oplus$ denote the exclusive OR, we
 consider the  
     binary WC defined by
     \beqn
     Y&=&X\oplus S \oplus \Psi, \label{eq:lkos-1}\\
     Z &=& X \oplus S \oplus\Phi,\label{eq:neko-chan-9}
     \eeqn
    where  $X, S, \Phi, \Psi$ are mutually independent. and $\Phi, \Psi$ play the role of {\em external} 
    "additive" noises 
    independent from the CSI $S$. We assume here that  $H(\Phi) > H(\Psi)$ and hence $Z$ is a degraded version of $Y$ in (\ref{eq:lkos-1}) 
    and (\ref{eq:neko-chan-9}). 
    \bteiri\label{teiri:RRR1} For the thus defined  binary degraded WC, we have
\beq\label{yatto-p}
C_{\mbox{{\scriptsize\rm CSI-E}}}  = C_{\mbox{{\scriptsize\rm NCSI-E}}} 
= H(\Phi) -H(\Psi).
\eeq
\eteiri
\bchui\label{chui:kimp2}
For comparison, let us consider the case where the encoder is {\em not} provided the CSI $S$.
In this case, it is natural to regard $S$ as an additive noise to the channel, then we have
 the secrecy capacity $C_{\mbox{\scriptsize M}}$ without CSI:

\beq\label{yatto-p1}
C_{\mbox{\scriptsize M}}
= H(S\oplus\Phi) -H(S\oplus\Psi).
\eeq
%
It is obvious that 
\beq\label{yatto-p2}
H(\Phi) -H(\Psi) > H(S\oplus\Phi) -H(S\oplus\Psi),
\eeq
which implies that  the existence of CSI $S$ can indeed outperform the channel without CSI.
Formula (\ref{yatto-p})  means that the secrecy capacity for this WC does not depend on $S$,
which is  a consequence of elimination of ``noise" $S$ by making use of  the CSI and is 
 in nice accordance with the formula of  Costa \cite{costa} on writing  on (Gaussian) dirty paper.
 A Gaussian counterpart is 
   discussed also  in 
Khisti {\em et al.} \cite{khisti-wornell}.
\IEEEQED
\echui

\smallskip
\noindent
{\it Proof of Theorem \ref{teiri:RRR1}:}  

Set  $X=U\oplus S$ where $U$ and  $X, S, \Phi, \Psi$ are independent, then 
     \beqn\label{eq:neko-chan-8}
     Y&=&U\oplus \Psi,\label{eq:lkos-2}\\
     Z &=& U \oplus \Phi.
     \eeqn
     }
     To show the achievability part, it suffices only to consider
     \beqn\label{eq:huro-1}
     \lefteqn{R_{\mbox{{\scriptsize\rm CSI-0}}}(p(u), p(x|u,s))}\non\\
     &=& I(U;Y)-I(U; Z)\non\\
     &=& H(U) - H(U|Y)- (H(U)-H(U|Z))\non\\
     &=& H(U|Z)-H(U|Y)\non\\
     &=& H(U|U\oplus \Phi) - H(U|U\oplus \Psi)\non\\
     &=& H(\Phi|U\oplus \Phi) - H(\Psi|U\oplus \Psi)\non\\
     &=& H(\Phi) -H(\Psi),
     \eeqn
     where the last step follows by setting $U\sim$(1/2, 1/2), which implies the achievability.
%
     %
   
   %
    On the other hand, in order to show the converse part, we invoke (\ref{eq:non-causal-conv-as2}) of
    Lemma \ref{hodai:state-enc-2}.
    %
    %
    %
    %
  %
    %
    %
  Let us evaluate the right-hand side of (\ref{eq:non-causal-conv-as2}) as follows:
  \beqn\label{eq:senegar-1}
 \lefteqn{I(U;Y)}\non\\
&=& I(U;X\oplus S\oplus \Psi)\non\\
 &\stackrel{(a)}{=}&I(U\oplus S; X\oplus \Psi)\non\\
  &=&I(X, U\oplus S; X\oplus \Psi)\non\\
  & & -I(X; X\oplus \Psi|U\oplus S)\non\\
   &=&I(X; X\oplus \Psi)\non\\
  & & -I(X; X\oplus \Psi|U\oplus S),
  \eeqn
  where in (a) we noticed that $(U, X\oplus S\oplus \Psi)$ and 
  $(U\oplus S, X\oplus \Psi)$ are in one-to-one correspondence under  operation
  $\oplus S$.

  Similarly, we have
    \beqn\label{eq:senegar-7}
 \lefteqn{I(U;Z)}\non\\
   &=&I(X; X\oplus \Phi)\non\\
  & & -I(X; X\oplus \Phi|U\oplus S).
  \eeqn
Hence,
\beqn\label{eq:senegar-9}
 \lefteqn{I(U;Y)-I(U;Z)}\label{eq:sosi-3}\non\\
 &=& I(X; X\oplus \Psi)\ -I(X; X\oplus \Phi)\non\\
 & & -(I(X;X\oplus \Psi|U\oplus S)-I(X; X\oplus \Phi|U\oplus S)).
 \eeqn
We now notice that $X\oplus \Phi$ is a degraded version of $X\oplus \Psi$ to obtain
\beq\label{eq:sosi-8}
I(X; X\oplus \Psi|U\oplus S)\ge I(X; X\oplus \Phi)|U\oplus S),
\eeq
from which together with (\ref{eq:senegar-9}) it follows that
\beq\label{eq:sosi-9}
I(U;Y)-I(U;Z) \le  I(X; X\oplus \Psi)\ -I(X;X\oplus \Phi).
\eeq
It is easy also
 to see that 
\beq\label{eq:sod-1}
\max_{p(x)} (I(X; X\oplus \Psi)\ -I(X;X\oplus \Phi)) = H(\Phi) -H(\Psi),
\eeq
where $\max$ can be attained with $X\sim (1/2.1/2)$, which implies  the converse.
\IEEEQED

\smallskip

In passing this section, let us look back at 
 Theorem \ref{teiri:RR-1} to scrutinize more the significance.
We first notice that the achievability of $R_{\mbox{{\scriptsize CSI-1}}}(p(u), p(x|u,s))$ in Theorem \ref{teiri:main1}
(and hence the achievability (\ref{eq:RR2}) in Theorem \ref{teiri:RR-1}) 
is  based on one-time pad cipher that  is attained by reproducing  CSI $S^n$ at Alice as $\hat{S}^n$ at Bob.
Furthermore, the achievability in Theorem \ref{teiri:bunin} with $V=(U, S)$ (and hence the achievability (\ref{eq:RR3})
in Theorem \ref{teiri:RR-1}) is also based on the reproduction of
  CSI $S^n$ at Alice as $\hat{S}^n$ at Bob as well.

In view of these observations along with  Remark \ref{chui:nebaneba-1}, 
we are now interested in what happens if we confine ourselves to 
within those coding schemes  that the CSI  $S^n$ at Alice  is required  to be reproduced as $\hat{S}^n$ at Bob
(this kind of coding schemes are said to be {\em state-reproducing}).
To see this, let the corresponding secret-message capacity and secret-key capacity be denoted by
the overlined quantities  as $\overline{C}$, then 
we have the following theorem:
\bteiri\label{teiri:Naga1}
For any degraded WC 
 ($Z$ is a degraded version of $Y$) with causal CSI only at Alice, we have
 \beqn
  \overline{C}_{\mbox{{\scriptsize\rm CSI-E}}} &=&  \overline{C}_{\mbox{{\scriptsize\rm NCSI-E}}} \nonumber\\
 &= & \max_{p(x|s)}\min \bigl(I(XS;Y) - I(XS;Z), I(XS;Y) -H(S)\bigr),
 \label{eq:RR2s}\\
 \overline{C}_{\mbox{{\scriptsize\rm CSI-E}}}^{\mbox{{\scriptsize\sf K}}}
&=&\overline{C}_{\mbox{{\scriptsize\rm NCSI-E}}}^{\mbox{{\scriptsize\sf K}}}\nonumber\\
 & =& \max_{\scriptsize\scriptsize\scriptsize\scriptsize 
 I(XS;Y) \ge H(S)}\bigl(I(XS;Y) - I(XS;Z) \bigr).
 \label{eq:RR3s}
  \eeqn
%
%
%
\eteiri
\bchui\label{chui:ma-chan-modoru}
It is easy to check that the right-hand side of (\ref{eq:RR2s}) is not greater than the right-hand side of (\ref{eq:RR3s}).
\IEEEQED
\echui
\smallskip

{\em Proof of (\ref{eq:RR2s}):} 

 It suffices to prove only  the converse. Since message $M$ and CSI $S^n$ are independent and 
 $S^n$ is reproducible at Bob, Fano inequality with  achievable rates
 $R$ and $\vep_n\to 0$ claims that
\beqn\label{eq:RR8s}
nR & =& H(M)\nonumber\\
&\le &H(M) - H(MS^n|Y^n) + n\vep_n\nonumber\\
&\le &H(MS^n) -H(S^n)- H(MS^n|Y^n) + n\vep_n\nonumber\\
&\le &I(MS^n; Y^n) -nH(S) + n\vep_n\nonumber\\
&=& \sum_{i=1}^nI(MS^n; Y_i|Y^{i-1})-nH(S)+n\vep_n\nonumber\\
&\le& \sum_{i=1}^nI(MS^nY^{i-1}; Y_i)-nH(S)+n\vep_n\nonumber\\
&\le& \sum_{i=1}^nI(X_iMS^nY^{i-1}; Y_i)-nH(S)+n\vep_n\nonumber\\
&=& \sum_{i=1}^nI(X_iS_iMS^{i-1}S_{i+1}^{n}Y^{i-1}; Y_i)-nH(S)+n\vep_n\nonumber\\
&\stackrel{(u)}{=}& \sum_{i=1}^nI(X_iS_i; Y_i)-nH(S)+n\vep_n\nonumber\\
   &=& \sum_{i=1}^n(I(X_iS_i; Y_i)-H(S_i))+n\vep_n\nonumber\\
 &\stackrel{(y)}{=}& n(I(XS; Y)-H(S))+n\vep_n,
\eeqn
%
where  
$(u)$ follows from the Markov chain property
$MS^{i-1}S_{i+1}^{n}Y^{i-1}\to X_iS_i\to Y_i$; 
 in $(y)$  we have used the argument similar to that in (\ref{eq:RR8}).
 Thus,  $R\le I(XS; Y)-H(S)$, which together with the proof of (\ref{eq:RR1}) implies the converse here.

{\em Proof of (\ref{eq:RR3s}):} 

 It suffices to prove only  the converse. Since  
 $S^n$ is reproducible at Bob, similarly to the derivation in (\ref{eq:RR8s}) we have
 \beqn\label{eq:nagata2}
 nH(S) &\le& H(S^n) -H(S^n|Y^n) +\vep_n\nonumber\\
 &=& I(S^n;Y^n)+\vep_n\nonumber\\
&=& \sum_{i=1}^n I(S^n;Y_i|Y^{i-1})+\vep_n\nonumber\\
&\le&  \sum_{i=1}^n I(S^nY^{i-1};Y_i)+\vep_n\nonumber\\
&\le& \sum_{i=1}^nI(X_iS_iS^{i-1}S_{i+1}^{n}Y^{i-1}; Y_i)+n\vep_n\nonumber\\
&=& \sum_{i=1}^nI(X_iS_i; Y_i)+n\vep_n\nonumber\\
 &=& nI(XS; Y)+n\vep_n.
 \eeqn
 Thus, $H(S)\le I(XS;Y)$, which together with the proof of (\ref{eq:RRM1})
implies the converse here.\IEEEQED

%
%
%
\section{Comparison with the previous result}\label{sec:previous}

We  have so far studied  the problem of how to convey confidential message over 
WCs with {\em causal} CSI available  only at Alice under the 
information leakage  $I_E=I(M^b; \sZ^b)\to  0$. 
In this connection, we notice  that this kind of problem with  {\em causal} CSI has not yet been 
brought to enough attention of  the researcher,
although the problem for WCs with {\em non-causal} CSI has extensively been investigated
in the literature.
On the other hand, to the best of our knowledge,  Fujita \cite{fujita} is supposed to be the first who has 
significantly addressed 
the problem  of  WCs with  {\em causal} CSI available only at Alice (used for key agreement with Bob),
although its {\em non-causal} counterpart had been studied by Khisti, Diggavi and Wornell \cite{khisti-wornell}.
 %
 %
 In this section,   we develop  the comparison with our results.

%
In order to describe   the main result of  \cite{fujita} in our terminology, define
 \beqn\label{eq:fuji-er}
 F_{\mbox{{\scriptsize CSI-1}}}(p(u), p(x|u,s))
& = & \min \Bigl[I(U; SY) -I(U; SZ) \non\\
& & \quad\qquad + H(S|Z) -H(S|Y),
 \nonumber\\
 & & \quad\qquad\qquad  I(U; SY)-H(S|Y)\Bigr],
 \eeqn
and let $C^{\mbox{w}}_{\mbox{{\scriptsize\rm CSI-E}}}$ denote the secrecy capacity under
the weak secrecy criterion $\frac{1}{n}I(M^b; \sZ^b) \to 0$ instead 
of $C_{\mbox{{\scriptsize\rm CSI-E}}}$.
Then, 

\medskip
\bteiri[\mbox{Fujita \cite[Lemma 1]{fujita}}]\label{hodai:neko-chan}
Let  us consider a degraded WC where $Z$ is a physically degraded version of $Y$, then
\beq\label{eq:fuji-san1}
C^{\mbox{w}}_{\mbox{{\scriptsize\rm CSI-E}}} \ge \max_{p(u), p(x|u,s)}
F_{\mbox{{\scriptsize\rm CSI-1}}}(p(u), p(x|u,s))
\eeq
holds.
\IEEEQED
\eteiri
\medskip

For comparison, 
we rewrite $F_{\mbox{{\scriptsize\rm CSI-1}}}(p(u), p(x|u,s))$ in (\ref{eq:fuji-er}) as follows.
\beqn\label{eq:fuji-er2}
 F_{\mbox{{\scriptsize CSI-1}}}(p(u), p(x|u,s))
& = & \min \Bigl[I(U; Y) -I(U; SZ) \non\\
& & \quad\qquad + H(S|Z) -H(S|UY),
 \nonumber\\
 & & \quad\qquad\qquad  I(U; Y)-H(S|UY)\Bigr],
 \eeqn
which is justified because
\beqn
I(U; SY)&=&I(U; Y)+I(U;S|Y),\\
H(S|Y)&=&H(S|UY)+I(U;S|Y).
\eeqn
Recalling that the lower bound $R_{\mbox{{\scriptsize CSI-1}}}(p(u), p(x|u,s))$ in Theorem \ref{teiri:main1} is
\beqn\label{eq:ogr-2}
R_{\mbox{{\scriptsize CSI-1}}}(p(u), p(x|u,s))
& = & \min \Bigl[I(U; Y) -I(U; SZ) \non\\
& & \quad\qquad + H(S|Z) -H(S|UY),\non\\
 & & \quad\qquad\qquad  I(U; Y)-H(S|UY)\Bigr]
\eeqn
and comparing it with (\ref{eq:fuji-er2}), it turns out that
$
R_{\mbox{{\scriptsize CSI-1}}}(p(u), p(x|u,s))
$
 exactly coincides with
$
F_{\mbox{{\scriptsize CSI-1}}}(p(u), p(x|u,s)).
$
Hence, the two largest lower bounds in Theorems \ref{teiri:main1} and  \ref{hodai:neko-chan} coincide with one another: 
\beq\label{eq:mer1}
\max_{p(u), p(x|u,s)}
R_{\mbox{{\scriptsize\rm CSI-1}}}(p(u), p(x|u,s))=\max_{p(u), p(x|u,s)}
F_{\mbox{{\scriptsize\rm CSI-1}}}(p(u), p(x|u,s)).
\eeq
On the other hand, the other largest lower bound in Theorem \ref{teiri:main1}:
\beq\label{eq:mer2}
\max_{p(u), p(x|u,s)}
R_{\mbox{{\scriptsize\rm CSI-2}}}(p(u), p(x|u,s))
\eeq
can be shown to be strictly larger than the left-hand side of (\ref{eq:mer1}) for an approximately selected  WC in which 
$Y$ is a degraded version of $Z$ (e.g., see \cite[Example 2]{chia-elgamal}), that is 
\beq\label{eq:mer3}
\max_{p(u), p(x|u,s)}
R_{\mbox{{\scriptsize\rm CSI-2}}}(p(u), p(x|u,s))>\max_{p(u), p(x|u,s)} 
R_{\mbox{{\scriptsize\rm CSI-1}}}(p(u), p(x|u,s)),
\eeq 
which together with (\ref{eq:mer1})  implies that for this WC the lower bound in Theorem \ref{teiri:main1} is strictly larger than 
the lower bound in Theorem \ref{hodai:neko-chan}.



%
%

Now, we are in a position to point out
 further crucial differences between \cite{fujita} and this paper,
 which is due to the completely different approaches taken to the problem.  
These are summarized as follows.

\begin{itemize}

\item
\cite{fujita} heavily depends on the assumption that the WC treated
needs to be  physically degraded, whereas this paper makes no such assumption.

\item
\cite{fujita} confines itself to within  the weak secrecy criterion problem ($\nth I(M; \sZ) \to 0$), whereas
this paper employs the strong secrecy criterion approach ($I(M; \sZ) \to 0$). As a consequence, all the results 
in \cite{fujita} (and \cite{chia-elgamal})  are guaranteed to hold as they are under the strong secrecy criterion too.

\item
In  \cite{fujita} all alphabets such as $\cU, \cS, \cX, \cY, \cZ$ are required to be finite, whereas in this paper 
$\cU,  \cX, \cY, \cZ$ except for $\cS$ may be arbitrary (including continuous alphabet cases), so that
Theorem \ref{teiri:main1} as stated  in Section \ref{intro-ge-cs0}  is directly applicable also  to, 
e.g.,   Gaussian 
WCs with causal CSI available   at Alice.

\item
The fundamental mathematical tool in  \cite{fujita} to deal with the problem is
the typical sequence argument (of course, well established), whereas in this paper 
the fundamental ingredients consist of Slepian-Wolf coding, Csisz\'ar-K\"orner's key construction, 
Gallager's maximum likelihood decoding, and 
Han-Verd\'u's resolvability argument (of course, well established).
This methodological difference brings about a new look  
from the viewpoint of   information theoretic perspective and 
applicability.
%
One of the consequences is that  the way of proving the main theorem here is significantly different from that in  \cite{fujita}.
This, for example, enabled us to naturally establish the strong secrecy property, which,
as is well known,  would not be quite easy to be attained 
by  the usual typical sequence arguments. 


\item
Most importantly, we see that there exists a crucial difference between \cite{fujita} and this paper
from the coding theoretic standpoint. Seemingly, both invoke the block Markov coding scheme as devised
 in \cite{cover-elgamal}, which is furnished with  a kind of  {\em forward-backward} 
 coding procedure. 

 \quad However,  in \cite{fujita}, the ``recursive" forward-backward coding procedure
 is employed in the sense that the $j$-th encoding in each block $j$ is carried out (which is carried over to
 the next block $j+1$) according to the order $j=1,2, \cdots, b.$
 During this procedure over the  total $b$ blocks no decoding is carried out.
When the encoding reaches the final block $j=b$ the decoding procedure gets started,
which is carried back to block $j=b-1$.
This  decoding  procedure is repeated backward according to the order $j=b, b-1, \cdots, 1$,
which causes at worst   ``$2b$ block decoding delay" in the whole process. 

\quad On the other hand, this paper employs the ``iterative" forward-backward coding procedure in the sense that
 not only the $j$-th encoding in each block $j$ (which is carried over to
 the next block $j+1$) but also the decoding for the previous block $j-1$ are carried out
 according to the order $j=1,2, \cdots, b.$ This  one-way coding scheme causes only ``one block decoding delay." 
%

\quad Why is this difference? The reason for this is that in \cite{fujita} the decoding operation in block $j$
is to be made upon receiving the information $\sS_j\sY_j$ but 
 the decoding operation for  $\sS_j$ is postponed  to the next block $j+1$ and it is in turn postponed  to block $j+2$,
 and recursively so on to reach the final block $j=b$. 
 Thus, actually, $\sS_j$ is decoded according to the order $j=b, b-1,  \cdots, 1$.
 In contrast with this, in this paper 
  the decoding operation in block $j$  is made upon receiving the information $\sY_j$,
 based on which $\sU_j$  is decoded and  used to decode $\sS_{j-1}$  in block $j-1$, and then proceed to 
 the next block $j+1$.
 This means that only  one block decoding delay and hence low complexities are needed.

\end{itemize}
%
%
%


%
%
%
%

%
\section{Concluding remarks}\label{conc-remark} 
In this last section, let us get started with quoting a paragraph from Chia and El Gamal \cite{chia-elgamal}, which addressed an interesting  non-trivial problem: 

“We used key generation from state information to improve the message transmission rate. It may be possible
to extend this idea to the case when the state information is available only at the encoder. This case, however,
is not straightforward to analyze since it would be necessary for the encoder to reveal some state information to the decoder (and, hence, partially to the eavesdropper) in order to agree on a secret key, which would reduce the
wiretap coding part of the rate.”

Motivated by it, we have investigated the coding problem for WCs with causal CSI at Alice and/or Bob, and established reasonable lower bounds on the secrecy capacity, which are summarized as Theorems 1 (one of the key results in this paper). Although Theorem 1 treats the WC with CSI available only at Alice, it can actually be useful enough for investigating general WCs with three correlated causal CSIs available at Alice, Bob and Eve, respectively. We would like to remind that this seemingly “general” WCs can actually be reduced to our WCs with CSI available only at Alice. In this connection, the reader may refer, for example, to Khisti, Diggavi and  Wornell \cite{khisti-wornell}, and Goldfeld, Cuff and Permuter \cite{goldfeld}.

As was pointed out in Section \ref{sec:previous},  the main ingredients thereby to establish Theorems \ref{teiri:main1}  actually consist of
the well established information-theoretic lemmas such as Slepian-Wolf coding, Csisz\'ar-K\"orner's key construction, Gallager’s maximum likelihood decoding, and Han-Verd\'u's resolvability argument, while not invoking the celebrated argument of typical sequences, which enabled us to well handle also the case with alphabets not necessarily finite, for example, including possibly  the case of Gaussian WCs with CSI.
Actually, this approach enabled us to derive some interesting results for degraded WCs as follows. Theorem 
\ref{teiri:RR-1} gives lower and upper bounds for the secret (-message) capacity, while, fortunately, the exact formula for the secret-key capacity has been determined. Corollary \ref{kei:han-sasaki} shows a causal secrecy capacity with one-sided CSI, which has nice correspondence with the interesting result of Chia and El Gamal with two-sided CSI \cite{chia-elgamal}, while 
Theorem \ref{teiri:RRR1} gives the secrecy capacity for binary WCs with one-sided CSI to establish a counterpart of Gaussian WCs studied by Costa \cite{costa} as “Writing on dirty paper.”
 Thus, these results together would provide a basic basis for  further investigation of  WCs with {\em causal}  CSI.

%
%
%
%
%


%

%
%
%
\appendices

\section{Proof of Lemma \ref{hodai:marchan}}\label{addA}
\noindent
{\em Proof of {\rm i)}: }
We can rewrite the security index
${\sf S}(\kappa(\sS_j)\sigma (\sS_j)|\sZ_j)$ in (\ref{eq:deketa-9}) of Lemma \ref{hodai:deketa-2} as
\beqn\label{eq:deketa-9s1}
\lefteqn{{\sf S}(\kappa(\sS_j)\sigma (\sS_j)|\sZ_j)}\non\\
 &=& {\sf S}(K_jN_{j+1}|\sZ_j)\non\\
 & = &D(P_{K_{j}N_{j+1}}||Q_{K_{j}N_{j+1}})+I(K_jN_{j+1}; \sZ_j)\non\\
 & \ge &I(K_jN_{j+1}; \sZ_j),
\eeqn
which together with Lemma \ref{hodai:deketa-2} gives i).\\
{\em Proof of {\rm ii)}: }
Here we use the following recurrence relation:
\beqn\label{eq:app-kor}
\lefteqn{I(K_{j-1}N_j; \sZ^{j-1}|\cH)}\non\\
&=& I(K_{j-1}N_j; \sZ_{j-1}|\cH)\non\\
& & +I(K_{j-1}N_j; \sZ^{j-2}|\sZ_{j-1}\cH)\non\\
&\le&  I(K_{j-1}N_j; \sZ_{j-1}|\cH)\non\\
& &+I(K_{j-2}N_{j-1}K_{j-1}N_j; \sZ^{j-2}|\sZ_{j-1}\cH)\non\\
&\stackrel{(j)}{=}&  I(K_{j-1}N_j; \sZ_{j-1}|\cH)\non\\
& &+I(K_{j-2}N_{j-1}; \sZ^{j-2}|\sZ_{j-1}\cH)\non\\
&\stackrel{(k)}{\le}&
 I(K_{j-1}N_j; \sZ_{j-1}|\cH)\non\\
& &+I(K_{j-2}N_{j-1}; \sZ^{j-2}|\cH),
\eeqn
where $(j)$ follows from the Markov chain property $ \sZ^{j-2} \to K_{j-2}N_{j-1} \to K_{j-1}N_j $
given $\sZ_{j-1}\cH$; $(k)$  follows from the Markov chain property $ \sZ^{j-2} \to K_{j-2}N_{j-1} \to  \sZ_{j-1}$
given $\cH$. Then, taking the summation of both sides in (\ref{eq:app-kor}) over $j\in [1: l]$ $(1\le l\le b)$ we have 
 \beqn\label{eq:app-kor1}
I(K_{l-1}N_l; \sZ^{l-1}|\cH) &\le& \sum_{j=1}^l  I(K_{j-1}N_j; \sZ_{j-1}|\cH)\non\\
& \le & \sum_{j=1}^b I(K_{j-1}N_j; \sZ_{j-1}|\cH)\non\\
&\stackrel{(m)}{=} & b\vep,
\eeqn 
where we have noticed that $I(K_{j-2}N_{j-1}; \sZ^{j-2}|\cH) =0$ for $j=1$
and $(m)$ follows from  i) of Lemma \ref{hodai:marchan}, thereby completing the proof.\IEEEQED

\section{Proof of Lemma \ref{hodai:deketa-5}}\label{addB}
The proof is carried out basically along the line of  Han and Verd\'u \cite[(8.3)]{ver-han} and Hayashi \cite[Theorem 3]{hayashi-exp}). We evaluate here the resolvability in terms of $I(M_{0j}; \sS_j\sZ_j|\cH)$
%
under rate constraint 
%
\beq\label{eq:app-1}
\overline{R}-R_0>I(U;SZ),
\eeq 
which is developed as follows. 

For each $m_0\in [1:2^{nR_0}],$  let $\sU(m_0)$ denote the random variable $\ssu_j(L(m_0))$
where $L(m_0)$ is distributed uniformly on the bin $\cB(m_0)$
 with rate constraint (\ref{eq:app-1}),
and 
define the channel 
\[
W(\sst|\ssu)\stackrel{\Delta}{=}P_{\sT(m_0)|\sU(m_0)}, 
\]
where $\sT(m_0) \stackrel{\Delta}{=} (\sS(m_0),\sZ(m_0))$, $\sst  \stackrel{\Delta}{=} (\sss, \ssz)$ and 
we notice that $P_{\sS(m_0)\sZ(m_0)|\sU(m_0)}$ does not depend on $m_0$, so that
we can write $P_{\sU\sS\sZ}$ instead of $P_{\sU(m_0)\sS(m_0)\sZ(m_0)}$. Now, set
\beq\label{eq:add-3}
L_n=2^{n(\overline{R}-R_0)}
\eeq
and
\beq\label{eq:add-5}
i_{\sU W}(\ssu, \sst)= \log\frac{W(\sst|\ssu)}{P_{\sT}(\sst)}.
\eeq
Then, 
\beqn\label{eq:add-6}
I(M_{0j};\sS_j\sZ_j|\cH)
&=& \frac{1}{2^{nR_0}}\sum_{m_0=1}^{2^{nR_0}}
\Exp_{\cH}D(P_{\sT(m_0)|\sU(m_0)}||P_{\sT(m_0)})\nonumber\\
& \stackrel{(a)}{=}&\Exp_{\cH}D(P_{\sT|\sU}||P_{\sT})\nonumber\\
& = &\sum_{\sst\in \cS^n\times \cZ^n}\sum_{\ssc_1\in\cU^n} \cdots  \sum_{\ssc_{L_n} \in \cU^n} P_{\sU}(\ssc_1)\cdots  P_{\sU}(\ssc_{L_n})\nonumber\\      
& &\quad \cdot \frac{1}{L_n}\sum_{j=1}^{L_n}W(\sst|\ssc_j)\log \left(\frac{1}{L_n} \sum_{k=1}^{L_n}\exp i_{\sU W}(\ssc_k, \sst)\right)\nonumber\\                                                                                
& = &\sum_{\ssc_1\in\cU^n} \cdots  \sum_{\ssc_{L_n} \in \cU^n} P_{\sU}(\ssc_1)\cdots  P_{\sU}(\ssc_{L_n})\nonumber\\      
& &\quad \cdot \sum_{\sst\in \cS^n\times \cZ^n}W(\sst|\ssc_1)\log \left(\frac{1}{L_n} \sum_{k=1}^{L_n}\exp i_{\sU W}(\ssc_k, \sst)\right)\nonumber\\   
& \stackrel{(b)}{\le} &\sum_{\ssc_1\in\cU^n}\sum_{\sst\in \cS^n\times \cZ^n}W(\sst|\ssc_1)P_{\sU}(\ssc_1)\nonumber\\     
& &\quad  \cdot \log \left(\! \frac{1}{L_n}\exp i_{\sU W} (\ssc_1, \sst) \mathalpha{+} \frac{1}{L_n}
\sum_{k=2}^{L_n}\Exp \exp i_{\sU W}(\sC_k, \sst)\!\right)\nonumber\\ 
& \stackrel{(c)}{\le} & \Exp \left[\log \left(1+ \frac{1}{L_n}\exp i_{\sU W}(\sU, \sT)\right)\right],
\eeqn
where $(a)$ follows from the symmetry of the random code $\cH$; $(b)$ follows from 
 the concavity of the logarithm; $(c)$  is the result of
\[
\Exp [\exp i_{\sU W}(\sC_k, \sst)] =1
\]
for all $\sst \in \cS^n\times\cZ^n$ and $k=1,2,\cdots, L_n$. Now, with $Q(\ssu)=P_{\sU}(\ssu)$,
apply a simple inequality 
with $0<\rho< 1$  and $x\ge 0$:
\[
\log (1+x) = \frac{\log (1+x)^{\rho}}{\rho}\le \frac{\log (1+x^{\rho})}{\rho}\le \frac{x^{\rho}}{\rho}
\]
to (\ref{eq:add-6}) to eventaully  obtain
\beqn\label{eq:add-7}
I(M_{0j}; \sS_j\sZ_j|\cH)
&\le &
\frac{1}{\rho L_n^{\rho}}\Exp \left(\frac{W(\sT|\sU)}{P_{\sT}(\sT)}\right)^{\rho}\nonumber\\
&=&\frac{1}{\rho L_n^{\rho}} \sum_{\sst\in \cS^n\times\cZ^n}\sum_{\ssu\in \cU^n}Q(\ssu)W(\sst|\ssu)
\left(\frac{W(\sst|\ssu)}{P_{\sT}(\sst)}\right)^{\rho}\nonumber\\
&=& \frac{1}{\rho L_n^{\rho}} \sum_{\sst\in \cS^n\times\cZ^n}\sum_{\ssu\in \cU^n}Q(\ssu)W(\sst|\ssu)^{1+\rho}
P_{\sT}(\sst)^{-\rho}.
\eeqn
On the other hand, by virtue of H\"older's inequality,
\beqn\label{eq:simple}
\lefteqn{\left(\sum_{\ssu\in \cU^n}Q(\ssu)W(\sst|\ssu)^{1+\rho}\right)P_{\sT}(\sst)^{-\rho}} \non \\
&=& \left(\sum_{\ssu\in \cU^n}Q(\ssu)W(\sst|\ssu)^{1+\rho}\right)
 \left(\sum_{\ssu\in \cU^n}Q(\ssu)W(\sst|\ssu)\right)^{-\rho}\non\\
&\le& \left(\sum_{\ssu\in \cU^n}Q(\ssu)W(\sst|\ssu)^{\frac{1}{1-\rho}}\right)^{1-\rho} 
\eeqn
for  $0 < \rho < 1$. Therefore, it follows from (\ref{eq:add-3}) that
\beqn\label{eq:app-8}
I(M_{0j}; \sS_j\sZ_j|\cH)
&\le & \frac{1}{\rho L_n^{\rho}} \sum_{\sst\in \cS^n\times\cZ^n}
 \left(\sum_{\ssu\in \cU^n}Q(\ssu)W(\sst|\ssu)^{\frac{1}{1-\rho}}\right)^{1-\rho} \non\\
 &=& \frac{1}{\rho}\exp\left[
 -[n\rho(\overline{R}-R_0) + E_0(\rho, Q)]\right],
\eeqn
where
\beq\label{eq:app-9}
E_0(\rho, Q) = -\log\left[
 \sum_{\sst\in \cS^n\times\cZ^n}
 \left(\sum_{\ssu\in \cU^n}Q(\ssu)W(\sst|\ssu)^{\frac{1}{1-\rho}}\right)^{1-\rho}
 \right].
 \eeq
Then, by means of Gallager \cite[Theorem 5.6.3]{gall},  we have $E_0(\rho, Q)|_{\rho=0}=0$ and 
\beqn\label{eq:app-10}
\left.\frac{\partial E_0(\rho, Q)}{\partial \rho}\right|_{\rho=0}
&=&
-I(Q, W)\non\\
&=&-I(\sU; \sS\sZ)\non\\
&\stackrel{(d)}{=}& -nIU;SZ), \non\\
\eeqn
where $(d)$ follows because $(\sU, \sS\sZ)$ is a correlated i.i.d. sequence with generic variable $(U, SZ)$.
Thus, for any small constant $\tau>0$ there exists a $\rho_0>0$ such that,
for all $0<\rho \le \rho_0$,
\beq\label{eq:app-11}
E_0(\rho, Q)\ge -n\rho(1+\tau)I(U;SZ)
\eeq
which is substituted into (\ref{eq:app-8}) to obtain
\beqn\label{eq:app-12}
\lefteqn{I(M_{0j}; \sS_j\sZ_j|\cH)}\non\\
&\le &  \frac{1}{\rho}\exp\left[
 -n\rho(\overline{R}-R_0 -(1+\tau) I(U;SZ))\right].
\eeqn
On the other hand, in view of  (\ref{eq:app-1}),  with some $\delta>0$ we can write 
\beq\label{eq:app-14}
\overline{R}-R_0=I(U;SZ)+2\delta,
\eeq
which leads to 
\beqn\label{eq:app-13}
\lefteqn{\overline{R}-R_0 -(1+\tau)I(U;SZ)}\non\\
&=& I(U;SZ) +2\delta
 - I(U;SZ) -\tau I(U;SZ)\non\\
&=& 2\delta -\tau I(U;SZ)).
\eeqn
We notice here that $\tau>0$ can be arbitrarily small,
so that the last term on the right-hand side of (\ref{eq:app-13})  can be made larger than $\delta>0$.
Then, (\ref{eq:app-12}) yields
\beq\label{eq:app-15}
I(M_{0j}; \sS_j\sZ_j|\cH) \le \frac{1}{\rho}\exp[-n\rho\delta],
\eeq
which implies that, for any small $\vep>0$, 
\beq\label{eq:app-15}
I(M_{0j}; \sS_j\sZ_j|\cH) \le \vep
\eeq
for all sufficiently large $n$, completing the proof of Lemma \ref{hodai:deketa-5}.
 \IEEEQED 

%
\section*{Acknowledgments}

The authors are grateful to Hiroyuki Endo for useful discussions. Special thanks go to Alex Bunin for useful comments,
which occasioned to improve Theorem \ref{teiri:main1}. 
Especially, the authors greatly appreciate the excellent editorial leadership of Matthieu Bloch 
who thoroughly read the earlier version to provide  insightful comments. 
This work  was funded by ImPACT Program of Council for Science, Technology and Innovation (Cabinet Office, Government of Japan).
\IEEEQED

%
%

%

\begin{thebibliography}{999}
%


\bibitem{wyner-wire} A. D. Wyner, ``The wire-tap channel," 
 {\em Bell Syst. Tech. J.}, vol.54, pp.1355-1387, 1975


\bibitem{csis-kor-3rd} I. Csisz\'ar and J. K\"orner, ``Broadcast channels with confidential messages,"
 {\em IEEE Transactions  Information Theory}, vol.24, no.3, pp.339-348, 1978
 
 \bibitem{slepian-wolf} D. Slepian and J.  K. Wolf, ``Noiseless coding of correlated information sources,"
 {\em IEEE Transactions  Information Theory}, vol.19,  pp.471-480, 1973

 
\bibitem{shannon-secrecy} C. E. Shannon, ``Communication theory of secrecy systems,"
 {\em  Bell Syst. Tech. J.},  vol. 28, pp.656-715, 1949
 
 \bibitem{shannon-cent} C. E. Shannon, ``Channels with side information at the transmitter,"
 {\em  IBM J. Tech. Develop.},  vol. 2, no. 4, pp.289-293, 1958

 
\bibitem{mitrpant}  C. Mitrpant, A. J. H. Vink and Y. Luo, ``An achievable region for the Gaussian wiretap channel with side information," {\em IEEE Transactions on Information Theory,} vol. 52, no. 5, pp. 2181-
2190, 2006

\bibitem{chen-vinck} Y. Chen and A. J. H. Vinck, ``Wiretap channel with side information,"
{\em IEEE International Symposium on Information Theory}, Seattle, USA, July 2006;
{\em IEEE Transactions on Information Theory,} vol. 54, no. 1, pp. 395-402, 2008


 \bibitem{liu-chen} W. Liu and B. Chen, ``Wiretap channel with two-sided channel state information."
 {\em 41st Asilomar Conference on Signals, Systems and Computation,} November, 2007
 
  \bibitem{bloch-lane} M. Bloch and J. N. Laneman, 
  ``Information-spectrum methods for Information-theoretic security,"
 {\em  Information Theory and Applications Workshop},  IEEE, 2009
 
 \bibitem{dai-vinck} B. Dai, Z. Zhuang and A. J. H. Vinck, ``Some new results on the wiretap
channel with causal side information,"{\em Proc. IEEE. ICCT,} Chengdu,
China, pp. 609-614, Nov. 2012.


\bibitem{boche} H. Boche and R. F. Schaefer, ``Wiretap channels with side information-
strong secrecy capacity and optimal transceiver design," {\em IEEE Transactions on information
Forensics and Security,} vol. 8, no. 8, pp. 1397-1408, 2013.

%
 \bibitem{maurer} U. M. Maurer, ``Secret-key agreement by public discussion from common information,"
 {\em IEEE Transactions on Information Theory,} vol. 39, no. 3, pp. 733-742, 1993
 
 %
 \bibitem{ahls-csis} R. Ahlswede and I. Csisz\'ar, 
 ``Common randomness in information theory and cryptography I,"
 {\em IEEE Transactions on Information Theory,} vol. 39, no. 4, pp. 1121-1132, 1993

 \bibitem{khisti} A. Khisti, S. Diggavi and G.Wornell, ``Secret-key agreement using asymmetry in
 channel state knowledge,"
 {\em  IEEE International Symposium on Information Theory}, Seoul, Proc.,  pp.2286-2290, 2009
 
  \bibitem{khisti-wornell} A. Khisti, S. Diggavi and G.Wornell, 
 ``Secret agreement with channel state information at the transmitter,"
{\em  IEEE Transactions on Information Forensics and Security}, 
no.3, vol.6, pp.672-681, 2011
 

 \bibitem{cover-elgamal} T. M. Cover and A. El Gamal, ``Capacity theorems for the relay channel,"
 {\em  IEEE Transactions on Information Theory}, vol.IT-25, no.5, pp.572-584, 1979
 
 \bibitem{chia-elgamal} Y.  K. Chia and A. El Gamal, ``Wiretap channel with causal state information,"
 {\em  IEEE Transactions on Information Theory}, vol.IT-50, no.5, pp.2838-2849, 2012
 
  
  \bibitem{dai-luo} B. Dai and Y. Luo, ``Some new results on the wiretap channel with side
information, {\em Entropy,} vol. 14, no. 9, pp. 1671-1702, 2012.


 \bibitem{sonee-hod} A. Sonee and G. A. Hodtani, ``Wiretap channel with strictly causal side
information at encoder," {\em Iran Workshop on Communication and Information Theory (IWCIT),}
2014



 \bibitem{fujita} H. Fujita,  ``On the secrecy capacity of wiretap channels with
side iinformation at the transmitter,"  {\em  IEEE Transactions on Information Forensics and 
Security}, vol.11, no.11, pp.2441-2452, 2016




\bibitem{hayashi-exp} M. Hayashi, ``Exponential decreasing rate of leaked information 
in universal random privacy amplification,"
 {\em  IEEE Transactions on Information Theory}, vol.IT-57, no.6, pp. 3989-4001, 2011
 
  \bibitem{han-et} T. S. Han, H. Endo and M. Sasaki, ``Wiretap channels with one-time state information: strong secrecy,"  {\em IEEE Transactions on Information Forensics and Security},
        vol.13, no.1, pp.224-236, 2018

 
 
 
 
  \bibitem{kramer} G. Kramer,  {\em Topics in Multi-User Information Theory,} Foundations and Trends in Communications and Information Theory, NOW, vol. 4: no. 45, pp 265-444. 2008
 
 \bibitem{prabhakaran} V. M. Prabhakaran, K. Eswaran and K. Ramchandran, ``Secrecy via Sources and Channels," {\em IEEE Transactions on  Information Theory,}  vol. 58, no.11, pp. 6747-6765,  2012



 
 
 
 \bibitem{goldfeld} Z. Goldfeld, P. Cuff, and H. H. Permuter,  
"Wiretap channel with random states non-causally available
at the encoder," https://arxiv.org/pdf/1608.00743v1. 2016

 
 \bibitem{bunin} A. Bunin, Z. Goldfeld, H. Permuter, S. Shamai, P. Cuff and P. Piantanida,"
 ``Semantically-secured message-key trade-off over wiretap channels with random parameters,"
https://arxiv.org/pdf/1708.04283, 2018; {\em Proc. of the 2nd Workshop on Communication Security}, pp. 33-48, 2018
  



 

 \bibitem{csis-kor-2nd} I. Csisz\'ar and J. K\"orner, {\em Information Theory: 
Coding Theorems for Discrete Memoryless Systems}, 2nd ed., Cambridge University Press, 2011
%

\bibitem{gall} R. G. Gallager, {\it Information Theory and Reliable Communication,}
John Wiley $\&$ Sons, NJ, 1968






\bibitem{gamal-kim} A. El Gamal and Y.H. Kim, {\it Network Information Theory,}
Cambridge University Press , New York, 2011%


\bibitem{hayashi-wire} M. Hayashi, ``General nonasymptotic and asymptotic formulas 
in channel resolvability and identification capacity and their application to the wiretap channel,"
 {\em  IEEE Transactions on Information Theory}, vol.IT-52, no.4, pp. 1562-1575, 2006
 


 
 


%
%

%
\bibitem{ver-han} T. S. Han and S. Verd\'u, ``Approximation theory of output statistics,"
 {\em  IEEE Transactions on Information Theory}, vol.IT-399, no.3, pp. 752-772, 1993
 %

\bibitem{costa} M.H. M. Costa, ``Writing on dirty paper," 
 {\em  IEEE Transactions on Information Theory}, vol.IT-29, no.3, pp. 439-441, 1983
 
 
\bibitem{han-sasaki-revisited} T. S. Han and M. Sasaki, ``Wiretap channels with causal state information: revisited," 
in preparation.

 
%
 %
%
 
%
 
%
%
 
%


 






 \end{thebibliography}
\end{document}